\documentclass[12pt, draftcls, onecolumn, letterpaper]{IEEEtran}
\ifCLASSINFOpdf
  
\else
 
\fi

\ifCLASSOPTIONcompsoc
    \usepackage[caption=false, font=normalsize, labelfont=sf, textfont=sf]{subfig}
\else
    \usepackage[caption=false, font=normalsize]{subfig}
\fi
\usepackage{lipsum}% 

\usepackage{balance}
\usepackage{multicol}   % for equalize of last page columns
\usepackage{cite}
\usepackage{gensymb}
\usepackage{multirow}
\usepackage{graphics}  
\usepackage{epsfig} 
\usepackage{graphicx}
\usepackage{epstopdf}
\usepackage{textcomp}
\usepackage{amsmath}
\usepackage{mathtools}
\usepackage{filecontents}
\usepackage{lipsum,color}
\usepackage{mathtools,tikz,caption}
\usepackage{amssymb}
\usepackage{amsfonts}
\usepackage{float,stfloats,blindtext}
\usepackage{hyperref}
\begin{document}
\title{Spatial Beam Tracking and Data Detection for an FSO Link to a UAV in the Presence of Hovering Fluctuations}

\author{Hossein~Safi,~Akbar~Dargahi,~and~Julian~Cheng,~{\it Senior Member, IEEE}}

%\author{Author A and Author B
%\author{Hossein~Safi,~Akbar~Dargahi~and~Majid~Safari
%\thanks{H. Safi and A. Dargahi are with the Department of Electrical Engineering, Shahid Beheshti University G. C., 1983963113, Tehran, Iran (e-mail: \{h$\_$safi, a$\-$dargahi\}@sbu.ac.ir). M. Safari is with the School of Engineering, the University of Edinburgh, EH8 9YL, Edinburgh, UK}
%}    

%% The paper headers
%\markboth{IEEE~PHOTONICS~TECHNOLOGY~LETTERS,~Vol.~XXX, NO.~XXX,~2018}
%{Safi \lowercase{\it et al}.: Joint Spatial Beam Tracking and Channel Estimation for Free-Space Optical Links.}

\maketitle

%%%%%%%%%%%%%%%%%%%%%%%%%%%%%%%%%%%%%%%%%%%%%%%%%%%%%%%%%%
%%%%%%%%%%%%%%%%%%%%%%%%%%%%%%%%%%%%%%%%%%%%%%%%%%%%%%%%%%
\begin{abstract}
	%%%%%%%%%%%%%%%%%%%%%%%%%%%%%%%%%%%%%%%%%%%%%%%%%%%%%%%%%%
	%%%%%%%%%%%%%%%%%%%%%%%%%%%%%%%%%%%%%%%%%%%%%%%%%%%%%%%%%%
	Recent advances in small-scale unmanned aerial vehicles (UAVs) have opened up new horizons for establishing UAV-based free-space optical (FSO) links. However, FSO technology requires precise beam alignment while random fluctuations of hovering UAVs can induce beam misalignment and angle-of-arrival (AoA) fluctuations. For an FSO link to a UAV, we consider a quadrant detector array for optical beam tracking and study the effect of random hovering fluctuations of the UAV on the performance of the tracking method, and based on the degree of instabilities for the UAV, the optimum size of the detectors for minimizing the tracking error is found. Furthermore, for optimal detection of On - Off keying symbols, the receiver requires instantaneous channel fading coefficients. We propose a blind method to estimate the channel coefficients, i.e., without using any pilot symbols, to increase link bandwidth efficiency. To evaluate the performance of the considered system, closed-form expressions of tracking error and bit-error rate are derived. Moreover, Monte-Carlo simulation is carried out to corroborate the accuracy of the derived analytical expressions.
%\\
%\\
%	\\
%	Due to
%	its ease of implementation, intensity modulation with direct detection (IM/DD) based on \textit{On--Off} keying (OOK)
%	is a popular signaling scheme in these systems. However, for optimal OOK detection the receiver
%	requires the instantaneous channel fading coefficients. Considering a quad-detector arrangement consisting of four avalanche
%	photo diodes, we propose a fast and practical method for spatial beam tracking and data detection. Particularly, we first determine the direction of arrival of the received optical beam at the receiver, and then estimate the channel blindly, i.e., without using any pilot symbols which increases the bandwidth efficiency of the link.  Therefore, OOK data detection is performed using the estimated channel coefficient. The proposed model
%	assumes an FSO link impaired by channel loss, atmospheric turbulence, and pointing errors. We provide extensive analysis of tracking error and bit error rate to evaluate the performance of the proposed method. The high accuracy of the analytical derivations is corroborated by
%	comparing numerically solved and Monte-Carlo simulation results.
\end{abstract}


\IEEEpeerreviewmaketitle
%%%%%%%%%%%%%%%%%%%%%%%%%%%%%%%%%%%%%%%%%%%%%%%%%%%%%%%%%%%%
%%%%%%%%%%%%%%%%%%%%%%%%%%%%%%%%%%%%%%%%%%%%%%%%%%%%%%%%%%%%
\vspace{-2.5 mm}
\section{Introduction}
\label{I}
\begin{figure}[t]
	\begin{center}
		\includegraphics[width=4 in]{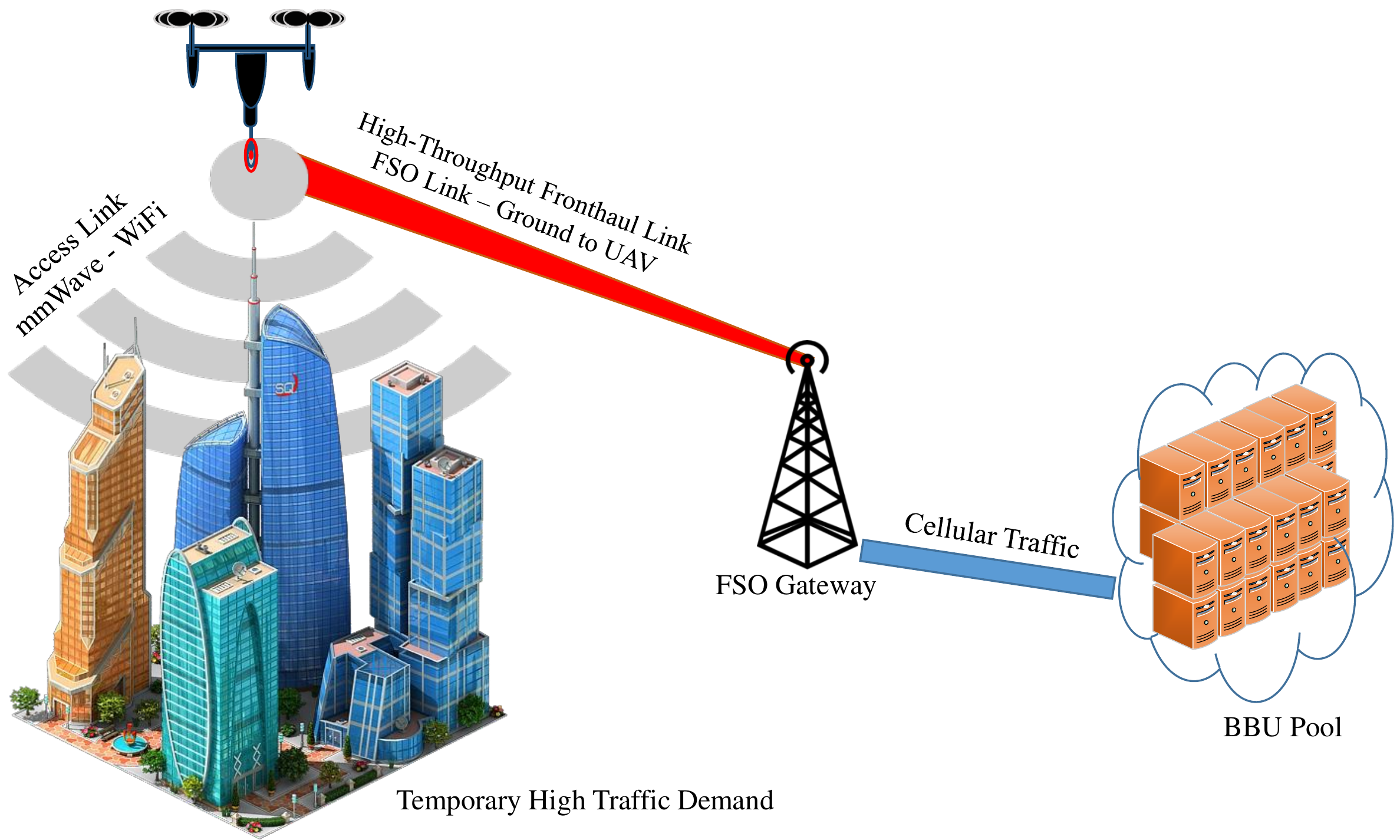}
		\caption{Graphical illustration of a ground-to-UAV FSO fronthaul link for 5G and beyond wireless networks.}
		\label{Ground-to-UAV}
	\end{center}
\end{figure}

%%%%%%%%%%%%%%%%%%%%%%%%%%%%%%%%%%%%%%%%%%%%%%%%%%%%%%%%%%%%
%%%%%%%%%%%%%%%%%%%%%%%%%%%%%%%%%%%%%%%%%%%%%%%%%%%%%%%%%%%%
\IEEEPARstart{R}{ecent} development of drone technology makes it possible to employ unmanned aerial vehicles (UAVs)
for wireless networking applications \cite{mozaffari2018tutorial}.  Inherent features of UAVs, such as mobility, flexibility,
and adaptive altitude adjustment allow fast and low-cost deployment of UAV communication networks compared
to their terrestrial counterparts. However, using radio frequency (RF)-based UAVs as aerial transceivers can cause interference to the existing terrestrial wireless networks. To avoid such radio interference as well as obtain high data rate transmission on the order of Gbps, as illustrated in Fig. \ref{Ground-to-UAV}, employing UAVs equipped with optical transceivers to establish free space optical (FSO)-based front-haul/back-haul links is proposed as a promising approach for the fifth-generation (5G) and beyond wireless networks \cite{alzenad2018fso,chen2017high,fawaz2018uav,dong2018edge}. However, FSO communication requires a stringent beam direction from transmitter to receiver. More precisely, random orientation deviations of hovering UAVs cause the fluctuations of angle-of-arrival (AoA) of optical beam at the lens aperture, which in turn cause image beam dancing at the photo-detector (PD) \cite{j2018channel}. Hence, it is essential to accurately aim the transmitter towards the receiver direction (pointing), and to determine the direction of arrival of the impinging beam at the receiver (spatial acquisition and tracking).

Acquisition, tracking, and pointing (ATP) for FSO links have been regarded as an interesting topic of research in the literature. Initial
studies addressed ATP in space optics, i.e., inter satellite links and earth--space laser communication. (see \cite[pp. 305-341]{gagliardi1995optical} and the references therein.) However, there exist fundamental differences between satellite-based and small UAV-based FSO systems, especially concerning the flying weight limit and AoA fluctuations owing to the orientation
deviations of hovering UAVs. Therefore, there is a need for in-depth investigation of the effective ATP mechanisms in this setup. 
%More recently,  ground-to-unmanned aerial vehicle (UAV) FSO links, as illustrated in Fig. \ref{Ground-to-UAV}, have been gathering momentum in this context  \cite{alzenad2018fso, chen2017high, 8570792,fawaz2018uav}.
% % % % % % % % % %
% % % % % % % % % %
% % % % % % % % % % % % %
% Inherent characteristics of UAVs such as mobility, flexibility, and adaptive altitude adjustment allow various types of
%them to be used in different situations, especially scenarios
%in which implementation of a UAV-based link, compared
%to its terrestrial counterparts, brings low cost and high
%speed implementation. However, there exist fundamental differences between satellite-based and small UAV-based FSO systems, especially, concerning the flying weight limit and angle of arrival (AoA) fluctuations owing to orientation
%deviations of hovering UAVs, which call for in-depth investigation of the effective ATP mechanisms in this setup. 
In a comprehensive literature review \cite{kaymak2018survey}, the ATP mechanisms for mobile FSO communications have been categorized according to their working principles, use cases, and used mechanics.  Nevertheless, due to the use of heavy and bulky mechanical or piezoelectric
equipments, e.g., gimbals and retro reflectors, most of them are not easily applicable for small-sized UAVs. More importantly, because of multi-gigabit transmission in FSO communication systems, beam tracking should be performed as fast as possible.

%However, most of these mechanisms need to employ beacon laser and also install heavy and bulky mechanical or piezoelectric equipments such as gimbal, servo motor, actuator, retro reflectors, etc, in the transceivers. Hence, they may not be adaptable to small- to medium-sized UAVs, vehicles, or mobile stations which are severely restricted by weight limitations. Moreover, due to high bit rate  transmission in FSO systems to provide ultimate fulfillment of 5G requirements, tracking approaches should be as fast as possible to compensate pointing errors quickly. 
%Adaptive beam divergence can also alleviate the problem of pointing errors \cite{kaymak2017divergence}. In particular, by wide transmitting of the optical beam the size of the beam footprint at the receiver increases results in  pointing error decreasing. Nevertheless, the wide divergence angle, the more geometric path loss for FSO systems.%For data transmission we employ intensity modulation with direct detection (IM/DD) based on \textit{{on}-{off}} keying (OOK) signaling scheme due to its ease of implementation. 
Using an array of PDs which are located at the focal plane of the receiver is another approach to perform optical beam tracking \cite{kiasaleh2015novel,kiasaleh2006beam, chalfant2017acquisition}. 
However, employing such arrangement of PDs gives rise to further challenges, e.g., the optimal size of the PDs and the essential of applying appropriate criterion for distinguishing between noise and received optical signal at the out put of the PDs. More precisely, optimizing the size of PDs involves reaching a compromise between the amount of undesired background power and mitigation of beam position deviation at the receiver. On the other hand, the challenge of determining the optimum criterion will be exacerbated when there is no knowledge of the channel state information (CSI) at the receiver side. 

%requires the knowledge of the optimal size of PDs to avoid accepting undesired background noise and minimize tracking error under different degrees of instabilities of UAVs. Second, the necessity of applying optimal criterion to determine signal and noise at the output of PDs and minimize tracking error.

%However, using such arrangement of photo detectors gives rise to further challenges, e.g., the optimal size of the photo detectors and the essential of applying appropriate criterion for distinguishing between noise and received optical signal at the out put of the photo detectors. More precisely, optimizing the size of photo detectors consists in reaching a compromise between the amount of undesired background power and mitigation of beam position deviation at the receiver. On the other hand, the challenge of determining the optimum criterion will be exacerbated when there is no knowledge of the channel state information (CSI) at the receiver side. 

%\textcolor{red}{Most of the above-mentioned works do not include tracking error analysis based on different channel impairments and consequently its effect on the system performance is neglected. Obviously, in order to design a reliable FSO communication system, the effect of spatial tracking error must be taken into account, which is dealt with in this paper. }

Due to the complexity of phase and frequency modulations and the associated implementation cost, intensity modulation with direct detection (IM/DD) based on \textit{{On}-{Off}} keying (OOK) signaling is widely adopted in the most current commercial FSO systems \cite{khalighi2014}.
In this signaling, bit `0' and bit `1' at each symbol interval are represented by the presence or absence of a light pulse, respectively. 
Compared to other signaling schemes, OOK also offers an improvement in bandwidth efficiency. However, for optimal data detection in this signaling, the receiver requires accurate knowledge of CSI to adaptively adjust detection  threshold under different channel fading conditions. Inevitably, prior to signal detection, the CSI should be accurately estimated at the receiver side. 
%Note that the channel fading in a FSO system is slowly varying, and for the typical data rates of FSO communication the channel coefficient remains unchanged over a large number of consecutive bits. 
Due to the inherent differences between optical systems and RF systems, especially regarding OOK signaling, power-dependent noise model, and avalanche photo detectors (APD)-based receivers, the classical RF channel estimation techniques are not appropriate for FSO communications. 
%Therefore, various data detection and channel estimation methods have been proposed in the context of FSO communications \cite{song2014robust,song2016robust,abadi2016fso,yang2015maximum,yang2016free,dabiri2017glrt,dabiri2018performance}.
More recently, sequence detection (SD) methods have been gathering strength in the context of FSO communications \cite{song2014robust,song2016robust,abadi2016fso,yang2015maximum,yang2016free,dabiri2017glrt}. In particular, by using SD methods, there is no need to estimate the channel via pilot symbols which leads to more bandwidth efficiency.  Moreover, to facilitate infrastructure transparency, it is far preferable to avoid data framing and packetization  at the transmitter \cite{4814379}. %The authors in \cite{song2014robust} have developed a Viterbi-type trellis-search sequence receiver based on the generalized likelihood ratio test (GLRT) principle that jointly detects the data sequence and estimates the unknown channel gain. In \cite{song2016robust} an efficient sequence receiver which acquires the CSI and the background information without knowledge about the channel model information has been proposed. A detection technique based on differential signaling scheme for  outdoor FSO communications is introduced in \cite{abadi2016fso}. Yang \textit{et al.} have applied maximum likelihood (ML) estimation to characterize the lognormal-Rician turbulence model parameters \cite{yang2015maximum}. Moreover, in \cite{yang2016free} source information transformation is investigated to detect an OOK signal without requiring the knowledge of instantaneous CSI. In our previous works, we have considered a GLRT-based sequence detection (SD) as an alternative to ML-based SD for the asymmetric OOK signaling \cite{dabiri2017glrt} and also proposed an expectation-maximization (EM)-based detection method to detect OOK symbols in a blind way \cite{dabiri2018performance}.
The aforementioned works have focused on two subjects: a) increasing bandwidth efficiency via detecting OOK symbols over the sequence of received signals in a blind way, i.e., without using any pilot bits, b) reducing the computational complexity of the proposed methods and making them fit to the maximum extent possible for high data rate FSO systems\footnote{Note that, the speed of opto-electronic devices is the main limiting factor to implement a high data rate FSO link. Hence, to increase electrical bandwidth and also to reduce the implementation cost, computational complexity should be decreased as much as possible.}. As mentioned earlier, in an FSO link, the receiver inevitably needs to track the orientation of the received optical beam before data detection. Therefore, to perform data detection and also spatial beam tracking, the receiver should blindly estimate the instantaneous channel fading coefficients. Also, in order to design a reliable FSO communication system, the effect of spatial tracking error must be taken into account, which is considered in this study.

% Most of the above-mentioned works do not include tracking error analysis based on different channel impairments and consequently its effect on the system performance is neglected. Obviously, in order to design a reliable FSO communication system, the effect of spatial tracking error must be taken into consideration.

%Note that, the speed of opto-electronic devices is the main limiting factor to implement a high data rate FSO link. 
%Regarding this, over the sequence of received optical signals and without requiring any pilot symbols, our aim is to design an FSO system that jointly track and detect data signals. 
%For spatial tracking, we consider a quad arrangement of avalanche photo detectors (APDs). For data transmission we employ IM/DD based on OOK signaling scheme.	
%In this paper, considering a quad arrangement of avalanche photo detectors (APDs) for spatial tracking.
%For data transmission we employ IM/DD based on OOK signaling scheme.

%Regarding this limitation, we propose a fast and practical method for  joint spatial tracking and data detection over impaired FSO links and then provide an extensive analysis of tracking error and bit error rate (BER) to evaluate the performance of the proposed method. Mathematical analysis and applied methodology are expressed in details and also closed-form formulations of BER and tracking error are derived. 

In this paper, considering a quad-detector arrangement consisting of four APDs and employing IM/DD based on OOK signaling scheme, for an FSO link to a UAV, we investigate the effect of hovering fluctuations on the performance of the tracking method and propose a fast and practical method for channel estimation and data detection. We address the above-mentioned challenges by employing a multi-element array of PDs at the receiver.  Specifically, we consider a practical scenario in which the receiver has no information about instantaneous channel fading coefficients. Therefore, over an observation window of length $L_s$ including several consecutive received OOK symbols, we first determine the direction of arrival of the impinging beam at the receiver and also estimate the channel state blindly. We then perform data detection using the results of the tracking step. 

{We investigate the effect of AoA fluctuations due to hovering fluctuations of UAVs on the performance of the proposed tracking method. As we will observe, hovering fluctuations severely deteriorate the performance of the tracking method. On the other hand, increasing the receiver field-of-view (FoV) via enlarging the size of photo detector can help  mitigate the performance degradation due to AoA fluctuations at the expense of accepting more background noise level and less electrical bandwidth of the receiver. Hence, we seek to find the optimum size for the employed quad-detector to minimize the tracking error under different degrees of hovering fluctuations of UAVs}. 

Moreover, we will show that the performance of the proposed methods for tracking and detection depends on the length of $L_s$ and  for an adequately large length of $L_s$ it acts like a receiver with known CSI. On the other hand, computational complexity, detection delay and also required memory increase linearly by increasing $L_s$. Hence, an optimum value of $L_s$ is not necessarily the biggest possible value, and optimizing $L_s$ deals with a trade-off between desired performance and the tolerable complexity of the system.
Since the target bit-error rate (BER) of FSO communication systems is usually lower than $10^{-9}$ \cite{navidpour2007ber,rockwell2001wavelength}, long processing time is required to carry out Monte-Carlo simulations. To overcome this time-consuming challenge, we provide an extensive analysis of tracking error and BER to evaluate the performance of the proposed methods. Mathematical analysis and applied methodologies are expressed in details. Also, closed-form formulations of BER and tracking error are derived. Simulation results verify the validity of the analytical derivations.
Regarding the practical purposes of establishing an FSO link to a UAV, our results can be used for easy calculating and tuning of the optimum value for the detector size as well as the length of observation window $L_s$ without employing Monte-Carlo simulations. 

%To the best of authors' knowledge, based on the open literature, there is no comprehensive work that perform tracking and detecting optical signals without requiring any pilot symbols.

%the proposed method is fast for  joint spatial tracking and data detection over impaired FSO links. and then provide an extensive analysis of tracking error and bit error rate (BER) to evaluate the performance of the proposed method. Mathematical analysis and applied methodology are expressed in details and also closed-form formulations of BER and tracking error are derived.

%, we investigate the effect of spatial tracking error on the system performance in a comprehensive approach. 

% To improve bandwidth efficiency via avoiding transmission of pilot symbols for channel estimation, a data detection method with implicit data-aided channel estimation have been proposed in this paper. Specifically, over an observation window of length $L_s$ encompassing several consecutive received OOK symbols, ...}

The rest of the paper is organized as follows. In Section
II, we describe the system model including the signal and channel model that will be used in this paper. In Section III, the spatial tracking and data detection methods are described followed by numerical results in Section IV. Finally, the paper is concluded in Section V.
%%%%%%%%%%%%%%%%%%%%%%%%%%%%%%%%%%%%%%%%%%%%%%%%%%%%%%%%%%%%
%%%%%%%%%%%%%%%%%%%%%%%%%%%%%%%%%%%%%%%%%%%%%%%%%%%%%%%%%%%%
%\vspace{-2 mm}
\section{System Model}
\label{II}
\begin{figure}
	\centering
	\subfloat[] {\includegraphics[width=4 in]{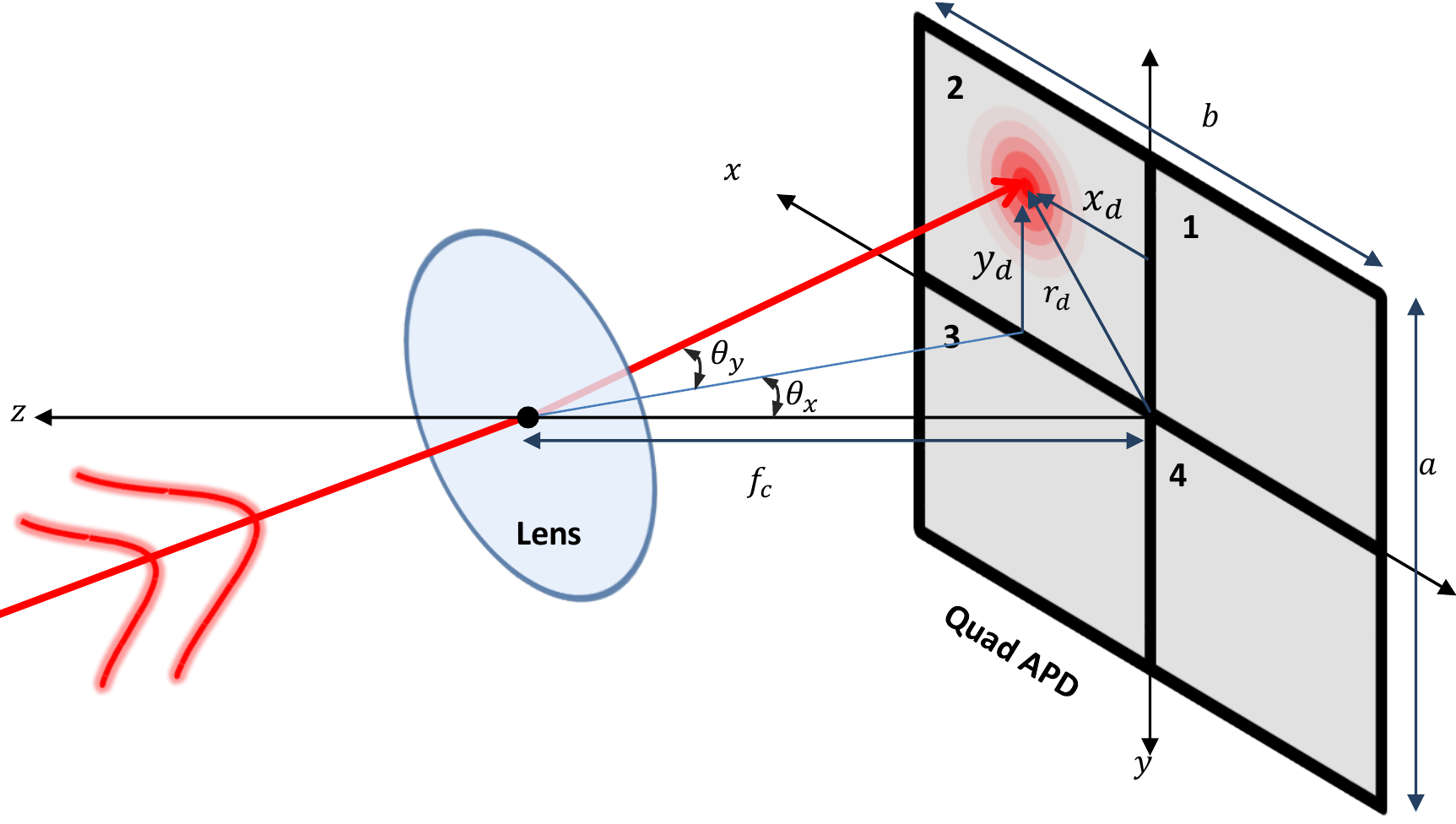}
		\label{M1}
	}
	\hfill
	\subfloat[] {\includegraphics[width=4 in]{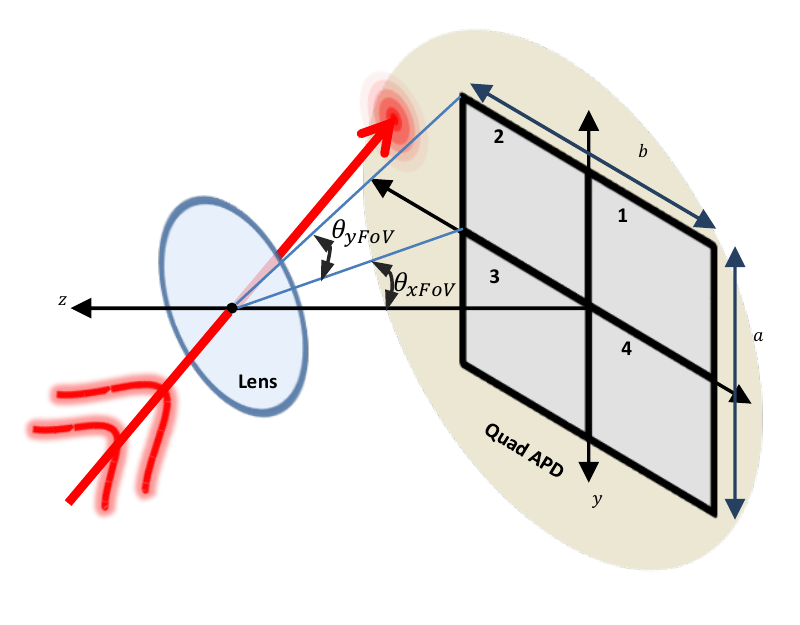}
		\label{M2}
	}
	\caption{{A schematic of the deviated received beam due to the pointing errors on the quad-APD detector when a) received laser beam is located on the quad-detector and b) full beam misalignment.}}
\label{b}
\end{figure}

{
%As shown in Fig. 1, a quad-APD detector with rectangular shape is employed in which the size area of the APDs are large enough to insure that the entire laser power through the aperture is focused onto the detector.
As shown in Fig. \ref{b}, a quad-APD detector with rectangular shape is employed.
We assume that the receiver aperture and the quad-detector are located on the $x-y$ plane and the beam propagated along $z$-axis. Let $\theta_x$ and $\theta_y$ denote the deviations of received laser beam due to the hovering fluctuations of UAV in $x-z$ and $y-z$ planes, respectively. The random variables (RVs) $\theta_x$ and $\theta_y$ are well modeled by the zero-mean Gaussian distribution with variance  $\sigma_x^2$ and $\sigma_y^2$, respectively, and their joint probability density function (PDF) is obtained as \cite{j2018channel}
\begin{align}
\label{orientation}
p_{\theta}\left( \theta_x,\theta_y\right) = \frac{1}{2\pi \sigma_x\sigma_y} \exp\left( -\frac{\theta_x^2}{\sigma_x^2}  -\frac{\theta_y^2}{\sigma_y^2}\right).
\end{align}
Moreover, we denote the receiver FoV in $x-z$ and $y-z$ planes by $\theta_{xFoV}$ and $\theta_{yFoV}$, respectively.
In this setup, we have $\theta_{xFoV}= {\rm arctan}\left(\frac{b}{f_c}\right)$ and $\theta_{yFoV}= {\rm arctan}\left(\frac{a}{f_c}\right)$ where $f_c$, $a$, and $b$ are the focal length of the lens and the detector's sides, respectively. Therefore, the receiver FoV in the spherical coordinate system can be represented by
}
{
\begin{align}
\label{sdfff}
\Phi_{FoV} &= 8 \int_0^{\tan^{-1}\left(\frac{a}{b}\right)}        \int_0^{\tan^{-1}\left(\frac{b}{2f_c \cos(\phi)}\right)}     \sin(\theta)d\theta d\phi  \\
%-------------------------------------------------
&= 8 \int_0^{\tan^{-1}\left(\frac{a}{b}\right)} 
\left[1 - \cos\left(\tan^{-1}\left(\frac{b}{2f_c \cos(\phi)}\right) \right) \right] d\phi. \nonumber
\end{align}
For the small values of $x$, via employing the small-angle approximation, we have $\cos(x)\simeq 1-\frac{x^2}{2}$ and also $\tan^{-1}(x)\simeq x$.
Since $a \,\&\, b << f_c$, eq. \eqref{sdfff} can be well approximated as
\begin{align}
\label{pd11}
\Phi_{FoV} &\simeq \frac{2 b^2}{f_c^2} \int_0^{\tan^{-1}\left(\frac{a}{b}\right)} 
\frac{1}{\cos^2(\phi)} d\phi 
\simeq \frac{2 ab}{f_c^2}.
\end{align}
}The entire received laser power through the aperture will be focused onto the detector if the  deviation of the received laser beam is smaller than the FoV of the receiver; otherwise, as shown in Fig. \ref{M2}, beam waist is placed out of the quad-detector which leads to the full optical beam misalignment.
\subsection{Signal Model}
%%%%%%%%%%%%%%%%%%%%%%%%%%%%%%%%%%%%%%%%%%%%%%%%%%%%%%%%%%%%
%%%%%%%%%%%%%%%%%%%%%%%%%%%%%%%%%%%%%%%%%%%%%%%%%%%%%%%%%%%%
%\begin{figure}
%	\begin{center}
%		\includegraphics[width=3.3 in]{Figures/Schematic.pdf}
%		\caption{A schematic drawing of the deviated received beam due to the pointing errors on the quad-APD detector}
%		\label{b}
%	\end{center}
%\end{figure}
We assume that IM/DD technique with OOK modulation is employed for signal transmission. Generally in practical FSO links, the mean of absorbed photons is sufficiently large; therefore, the distribution of the number of APD output electrons can be well approximated by Gaussian \cite{davidson1988gaussian}. Thus, the photo-current corresponding to the $k$-th symbol interval and  the $i$-th quadrant of the quad-detector can be expressed as
\begin{equation}
\label{photocurrent}
r_{i}[k] =  h D_i \mu s[k] + n_i[k], \text{~~~for~~~} i \in \{1, \ldots, 4\}
\end{equation}
where $h$ is the channel coefficient including the
channel loss, the effects of atmospheric turbulence, and pointing errors and is also assumed to be constant over a large sequence of transmitted bits (i.e., slow fading channel). The area of the photo detectors is assumed to be larger than the beam waist, hence, with good accuracy when the beam deviation is smaller than the receiver FoV, it can be assumed that at each interval the deviated received beam is focused onto the $i$-th quadrant of the quad-detector. { Note that, the fraction of power in side lobes of Airy pattern is much smaller than that in main lobe and ignoring the effect of power in side lobes of Airy pattern can be a reasonable assumption \cite{gagliardi1995optical,kiasaleh2006beam}. Also, the width of the main lobe of the Airy pattern is approximately equal to $2.4\lambda$ and it is much smaller than the conventional size of an APD which is commonly in order of \rm{mm} \cite{gagliardi1995optical}. Hence, we ignore the effect of boundary conditions of the main lobe. } 

In \ref{photocurrent}, the parameter $D_i \in \{0,~1\}$ indicates the presence of the received beam at the $i$-th quadrant. Accordingly, {when the received laser beam is placed at the receiver FoV,} at each transmission interval, $D_i = 1$ implies that only the $i$-th quadrant of the quad-detector captures the received laser beam and the remaining three quadrants do not receive transmitted optical signal.
Moreover, under the condition of full beam misalignment, we have $D_i=0$ for $i\in\{1, ..., 4\}$.
Due to the Gaussian distributed RVs $\theta_x$ and $\theta_y$, the probability of having full beam misalignment can be obtained as
%%%%%%%%%%%%%%%%%
\begin{align}
\label{interrupt}
\mathbb{P}_{fbm}= 1- \sum_{i=1}^4 \mathbb{P}_{D_i}
\end{align}
where $\mathbb{P}_{D_i}$ is the probability of capturing the arrival beam at the $i$th quadrant. Because of the symmetry of the quad-detector arrangement, we have  
$\mathbb{P}_{D_i}= \mathbb{P}_{D_j}=\mathbb{P}_{D}$ for $i\,\&\,j\in\{1,...,4\}$, and for instance $\mathbb{P}_{D_1}$ can be obtained as
{
\begin{align}
\label{pd1d}
\mathbb{P}_{D_1} &= \int_0^{\tan^{-1}\left(\frac{a}{f_c}\right)} \int_0^{\tan^{-1}\left(\frac{b}{f_c}\right)}  p_{\theta}\left( \theta_x,\theta_y\right)  d\theta_xd\theta_y  \\
&=\left(\frac{1}{2}-Q\left( \frac{\tan^{-1}\left(a/f_c\right)}{\sigma_x}\right) \right)
  \left(\frac{1}{2}-Q\left( \frac{\tan^{-1}\left(b/f_c\right)}{\sigma_y}\right) \right). \nonumber
\end{align}
}Moreover in (\ref{photocurrent}), $\mu= \frac{e G \eta}{h_p \nu}$  where $e$ denotes the charge of electron, $G$ is the average APD gain, $\eta$ denotes the APD quantum efficiency, $h_p$ denotes the Planck\textquotesingle s constant, and $\nu$ is the optical frequency. Furthermore, $s[k]$ and $n_i[k]$, respectively, stand for the transmitted symbol with optical power $P_t$ and the photo-current noise of the $i$-th quadrant which is an additive white Gaussian noise (AWGN) with zero-mean and variance $\sigma_{i,k}^2$ which is given by
\begin{align}
\label{variance-noise-i,k}
\sigma_{i,k}^2 = \sigma_s^2 h D_i s[k] + \sigma_0^2, \text{~~~for~~~} i \in \{1, \ldots, 4\}
\end{align}
where $\sigma_s^2$ is the variance of the shot noise due to transmitted signal and $\sigma_0^2 = \sigma_{b}^{2} + \sigma_{th}^{2}$ is the total noise variance due to the variance of background radiation, $\sigma_{b}^{2}$, and receiver thermal noise, $\sigma_{th}^{2}$. We have $\sigma_{b}^{2} = 2 e G F \mu B P_b$ where $F$ denotes the APD excess noise factor, and $B$ is the bandwidth of the receiver low-pass filter (in Hz). Furthermore, $\sigma_{th}^{2}=\frac{4K_{B}T_{r}B}{R_{l}}$, where $ K_{B} $ is Boltzmann constant, $ T_{r} $ is the receiver equivalent temperature in Kelvin, and $ R_{l} $ is the load resistance. 
The background power $P_b$ is a function of the photo-detector area and can be attained as \cite{j2018channel}
\begin{align}
\label{xf1}
P_b = N_b(\lambda)\, B_o\, \Omega_{FoV}\, A_a
\end{align}
where $N_b(\lambda)$ is the spectral radiance of the background radiations at wavelength $\lambda$ (in Watts/${\rm cm}^2$-$\micro$m-srad), $B_o$ is the bandwidth of the optical filter at the Rx (in $\micro$m), and $A_a$ is the lens area (in ${\rm cm}^2$). Regarding  (\ref{pd11}), in our setup, $P_b$ is derived as
\begin{align}
\label{Pbb}
P_b =   \frac{2 a\,b\, N_b(\lambda)\, B_o\,A_a}{f_c^2}. 
\end{align}
%Explicitly, $\sigma_n^2$ incorporates photo-current noise
%The received signal $r_k$, can be expressed at any discrete time $  k$ as
%$r_k = P_t s_kh + n_k$,
%where $ h $ is the fading channel coefficient \textcolor[rgb]{1,0,0}{that is assumed constant over a large number of transmitted bits,} $ s_k\in \{\alpha_e,1\} $ is the transmitted NRZ-OOK symbol, $s_k=\alpha_e$ represents a digital symbol $s_{k}'=0$ while $s_k=1$ represents a digital symbol $s_{k}'=1$ \cite{ghassemlooy_book}. Without loss of generality, we assume that the transmitted energy $P_t$ is normalized to one. Also, we assume that $n_k$ is the additive white Gaussian noise with zero mean and variance $\sigma_1^2$ due to the transmitted one-bit and $\sigma_0^2$ due to the transmitted zero-bit \cite{moradi2010, holzman_JLT}. Moreover, the average electrical SNR is defined as \textcolor[rgb]{1,0,0}{$\frac{2E\{|s_kh|^2\}}{\sigma_1^2+\sigma_0^2} $,} where $ E\{.\} $ denotes statistical expectation. 
%In order to detect FSO signals, the received data is gathered in an observation window of  $ L_{s} $ interval, $\left[r_k\right]_{k=1}^{L_{s}}$, corresponding to  the $ L_{s} $ transmitted signals, $\left[s_k\right]_{k=1}^{L_{s}}$ during which the channel is assumed to constant.
%%%%%%%%%%%%%%%%%%%%%%%%%%%%%%%%%%%%%%%%%%%%%%%%%%%%%%%%%%%%
%%%%%%%%%%%%%%%%%%%%%%%%%%%%%%%%%%%%%%%%%%%%%%%%%%%%%%%%%%%%
\subsection{FSO Channel Model}
%%%%%%%%%%%%%%%%%%%%%%%%%%%%%%%%%%%%%%%%%%%%%%%%%%%%%%%%%%%%
For channel modeling we consider three impairments, namely, the deterministic propagation loss $h_{loss}$, the atmospheric turbulence $h_{atm}$, and the pointing error loss $h_{poi}$. Therefore, the channel coefficient $h$ is represented by
\begin{align}
\label{channel-coefficient}
h = h_l h_{atm} h_{poi}.
\end{align} 
Assuming Gamma-Gamma atmospheric turbulence channels, the PDF of $ h $ is given by \cite[eq. (12)]{sandalidis2009optical}
\begin{align}
\label{fading}
f_{h}(h)=  \frac{\alpha\beta\gamma^{2}}{A_{0}h_{l}\Gamma(\alpha)\Gamma(\beta)}         
	\times G_{1,3}^{3,0}   \left( \frac{\alpha\beta}{A_{0}h_{l}}h  \  \Bigg\vert \  {\gamma^{2} \atop \gamma^{2}-1,\alpha-1,\beta-1} \right)
\end{align}
where $  G_{1,3}^{3,0}   \left( \cdot \right) $ is the Meijer\textquotesingle  s $  G$ function, and $\Gamma(\cdot)$ is the Gamma function. The parameter $ \gamma$=$w_{L_{eq}}/2\sigma_{j} $   denotes the ratio between the equivalent beam radius at the receiver and the pointing errors displacement standard deviation. Furthermore,  $w_{L_{eq}}^{2}$=$w_{L}^{2}\sqrt{\pi} .{\rm erf}(v)/\left(2v\exp\left (-v^{2}\right )\right )   $, in which $ w_{L} $ is the beam radius at the distance $  d_{0}$, ${\rm erf(\cdot)}$ is the error function,
$ v$=$\sqrt{\pi}r/\left (\sqrt{2}w_{L}\right ) $, and  $r$ is the radius of a circular detector aperture. Also, the parameter $A_{0}$=$[{\rm erf}(v)]^{2}$ denotes the maximal fraction of the collected power. Furthermore, $  1/\beta$ and $1/\alpha$ are, respectively, the variances of the small scale and large scale eddies which are given by
\begin{equation}
\label{alpha}
1/\alpha = \left[     \exp     \left(      \frac{0.49\chi^{2}}      {\left(1+1.11\chi^{12/5}  \right)^{7/6}}\    \right) -1    \right]
\end{equation}
and
\begin{equation}
\label{betta}
1/\beta = \left[     \exp     \left(      \frac{0.51\chi^{2}}      {\left(1+0.69\chi^{12/5}  \right)^{5/6}}\    \right) -1    \right]
\end{equation}
where $ \chi^{2} $ is the Rytov variance. Indeed, for a slant path between transceivers, $\chi^{2}$ can
be expressed as a function of the link length $L$, and the height difference between transceivers, $x_r$, as \cite{laserbook}
\begin{equation}
\label{dfg2}
\chi _i^2(L,x_r) = 2.25\left(\dfrac{2\pi}{\lambda}\right)^{\frac{7}{6}}
\left(\frac{L}{x_r}\right)^{\frac{11}{6}}
\int_{0}^{x_r} C_n^2(x)\left( x - \frac{x^2}{d_v}\right)^{\frac{5}{6}} dx,
\end{equation}
where $C_n^2(x)$ is the refractive-index structure parameter based on Hufnagel-Valley (HV) model, and is expressed as \cite{laserbook}
%\begin{equation}
\begin{align}
\label{eq-5}
C_n^2(x) =&~ 0.00594(\mathcal{V}/27)^2(10^{-5}x)^{10}\exp(-x/1000) \nonumber \\
&+ 2.7\times10^{-16}\exp(-x/1500) + C_n^2(0)\exp(-x/100)
\end{align}
%\end{equation}
where $\mathcal{V}$ is the speed of strong wind, and $C_n^2(0)$ is the nominal value of refractive-index structure parameter at ground level in $m^{-2/3}$.
\section{Spatial Tracking and Data Detection}
\label{III}
For an observation window of length $L_s$ bits, the received signal vector $\underline{r_i} = \{r_{i}[1],r_{i}[2],\ldots,r_{i}[L_s]\}$ at the $i$-th quadrant of the quad-detector is related to the $L_s$ transmitted signal vector $\underline{s} = \{s[1],s[2],\ldots,s[{L_s}]\}$. We also assume slow fading channel, i.e., channel remains constant during observation window. Note that for performing optical beam tracking as well as OOK demodulation the knowledge of the CSI should be available with pinpoint accuracy at the receiver. In the sequel, we first study spatial beam tracking under the assumptions of known CSI at the receiver. However, for unknown CSI scenarios, we propose an efficient data-aided channel estimation method without inserting any pilot symbol encounters a signaling overhead. We then investigate the spatial tracking problem based on the estimated channel and evaluate the performance of the proposed approach. Further, the method that we have adopted for data detection will be introduced in the second part of this section.
%%%%%%%%%%%%%%%%%%%%%%%%%%%%
%%%%%%%%%%%%%%%%%%%%%%%%%%%%%
%%%%%%%%%%%%%%%%%%%%%%%%%
\subsection{Spatial Tracking}
%%%%%%%%%%%%%%%%%%%%%%%%
Let us denote $m$ as the number of bits `1' in the observation window of length $L_s$, i.e., $m = \sum\limits_{k = 1}^{{L_s}} {s[k]}$.  %Moreover, $m$ is unknown to the receiver. 
Accordingly, at the $i$-th quadrant, the received signal (photo-current) conditioned on $h$ and $m$ can be written as 
\begin{align}
\label{conditioned-r}
{r'_{i|h,m}} = \sum\limits_{k = 1}^{{L_s}} {{r_i}[k] = h{D_i}} \mu m + {n'_{i|h,m,{D_i}}}
\end{align}
where ${n'_{i|h,m,{D_i}}} = \sum\limits_{k = 1}^{{L_s}} {n_{i}[k]}$ is an AWGN with zero mean and variance as follows
\begin{align}
\label{variance-noise-i}
\sigma_{i|h,m,D_i}^2 = \sigma_s^2 h D_i m + L_s\sigma_0^2.
\end{align}                     
Hence, the PDF of ${r'_{i|h,m}}$ conditioned on $D_i$ can be obtained as
\begin{align}
\label{pdf-r-conditional}
p(r'_{i|h,m}|{D_i}) \!=\! \frac{1}{{\sqrt {2\pi \sigma _{i|h,m,{D_i}}^2} }}\exp \!\!\left( - {\frac{{{{\left| {r{'_{i|h,m}} - h{D_i}\mu m} \right|}^2}}}{{2\sigma _{i|h,m,{D_i}}^2}}} \right)\!\!.
\end{align}
In the following, we proceed to perform spatial tracking when the CSI is either known or unknown but estimated at the receiver side.
%%%%%%%%%%%%%%%%%%%%%%%%%
\subsubsection{Known CSI}
%%%%%%%%%%%%%%%%%%%%%%%%
When $h$ is known, the receiver decides that the laser beam is captured by the $i$-th quadrant based on an maximum likelihood (ML) criterion which is expressed as
\begin{align}
\label{ML-criterion}
\hat i =& \underset{i \in \{1,\ldots,4\}}{\operatorname{\text{~arg~max~}}}~p(r'_{i|h,m}|{D_i=1})\times \!\!\!\!\!\!\prod_{j=1,j\ne i}^4 \!\!\!\!p(r'_{j|h,m}|{D_j=0}), \nonumber \\ 
=& \underset{i \in \{1,\ldots,4\}}{\operatorname{\text{~arg~max~}}} \log\left(p(r'_{i|h,m}|{D_i=1})\right)+ \!\!\!\!\!\sum_{j=1,j\ne i}^4 \!\!\!\!p(r'_{j|h,m}|{D_j=0}). 
\end{align}
Substituting (\ref{pdf-r-conditional}) into (\ref{ML-criterion}) and after some manipulations, the spatial beam tracking based on metric $\mathcal{M}_{i|h,m}$ can be stated as follows
\begin{align}
\label{metric-tracking-1}
\hat i = \underset{i \in \{1,\ldots,4\}}{\operatorname{\text{~arg~min~}}}~\mathcal{M}_{i|h,m}
\end{align}
where
\begin{align}
\label{metric-tracking-2}
\mathcal{M}_{i|h,m} = \frac{{{{\left| {{{r'}_{i|h,m}} - h\mu m} \right|}^2}}}{{\sigma _s^2hm + {L_s}\sigma _0^2}} +\!\!\! \sum_{j=1,j\ne i}^4 \!\!{\frac{{{{\left| {{{r'}_{j|h,m}}} \right|}^2}}}{{{L_s}\sigma _0^2}}}. 
\end{align}
Tracking error probability of this method is derived in Appendix A as 
\begin{align}
\label{TER_PCSI}
&\mathbb{P}^{\rm p}_{te} \simeq \mathbb{P}_{fbm} +  \frac{(1-\mathbb{P}_{fbm})}{2^{L_s}}  \int_0^\infty \sum_{m=0}^{L_s} \binom{L_s}{m} \Bigg\{1- \nonumber \\
&\left( 1-Q\left(  \frac{\mu \sigma_s^2 h^2m^2 \left(h\mu m+2L_s\sigma_0^2\right)}
{\sigma_{tc|h,m}}  \right) \right)^3 \Bigg\}f_h(h)dh.
\end{align}
where
\begin{align}
\label{hjj3}
\sigma_{tc|h,m}^2  =&~\left(2\sigma_s^2 h m \left(h m + L_s\sigma_0^2 \right)\right)^2 
\times \left(\sigma_s^2 h m + L_s\sigma_0^2\right)  \nonumber \\ 
&~+L_s\sigma_0^2   \left(2mL_s\sigma_0^2 \sigma_s^2 h\right)^2  .
\end{align}

%%%%%%%%%%%%%%%%%%%%%%%%%%%%%%%%%%%%%%
\subsubsection{Unknown CSI}
%%%%%%%%%%%%%%%%%%%%%%%%%%%%%%%%%%%%%%
In this part, we consider a scenario in which the parameters $m$ and $h$ are not known at the receiver. Thus, we have to modify the proposed metric in \eqref{metric-tracking-1} based on this practical assumption. At first step, we have to estimate $h$ from the received data sequence over the observation window of length $L_s$. When the received laser beam is placed at the receiver FoV, at each transmission interval, laser power is focused onto one quadrant of the quad-detector, i.e., $\sum\limits_{i = 1}^4 {{D_i}}=1$. 
Therefore, the total photo-current $r[k]$, generated by the quad-detector can be obtained as
%We use this point and define $r[k]$ as
\begin{align}
\label{xd1}
r[k] = \sum\limits_{i = 1}^{{4}} {r_{i}[k]}  = h \mu s[k] + n[k]
\end{align}
where $n[k] = \sum\limits_{i = 1}^{{4}} {n_{i}[k]}$ is an AWGN with zero-mean and variance
$\sigma_{k}^2 = \sigma_s^2 h s[k] + 4\sigma_0^2$.
From \eqref{xd1}, during the observation window, an estimation of $h$ can be obtained as
\begin{align}
\label{xd2}
\hat{h}  = \dfrac{2}{\mu L_s}\sum_{k=1}^{L_s} r[k] = \frac{2}{L_s}R.
\end{align}
Substituting \eqref{xd1} into \eqref{xd2}, we have
\begin{align}
\label{xd3}
\hat{h}  = \frac{2m}{L_s}h + n_{\hat{h}}
\end{align}
where $n_{\hat{h}} = \frac{2}{\mu L_s}\sum_{k=1}^{L_s} n[k]$, and is a Gaussian  noise with zero-mean and variance 
\begin{align}
\label{kx5}
\sigma_{h}^2 = \frac{4}{(\mu L_s)^2}\left(m\sigma_s^2 h + 4L_s\sigma_0^2\right).
\end{align}
Clearly, one can see from \eqref{kx5} that $\sigma_{h}^2$ will tend to zero by increasing $L_s$.  
On the other hand, it can be assumed that for large values of $L_s$, the number of `1's in the observation window of length $L_s$ is likely to be close to its expected value, i.e., $m \approx \dfrac{L_s}{2}$.
Therefore, by using $\frac{L_s}{2}$ and $\hat{h}$ obtained from \eqref{xd3} instead of $m$ and $h$ in the decision rule \eqref{metric-tracking-1}, the modified proposed metric under unknown CSI condition can be stated as 
\begin{align}
\label{xd6}
\hat i = \underset{i \in \{1,\ldots,4\}}{\operatorname{\text{~arg~min~}}}~\mathcal{M}'_{i|h,m}
\end{align}
where
\begin{align}
\label{xd7}
\mathcal{M}'_{i|h,m} = \frac{{{\left| r'_{i|h,m} - \mu R \right|}^2}}{{\sigma _s^2R + {L_s}\sigma _0^2}} +\!\!\!\!\!\! 
\sum_{j=1,j\ne i}^4 \!\!\!\!\!{\frac{{{{\left| {{{r'}_{j|h,m}}} \right|}^2}}}{{{L_s}\sigma _0^2}}}. 
\end{align}
Tracking error probability of the proposed method under unknown CSI is derived in Appendix B as 
\begin{align}
\label{TER_ICSI}
&\mathbb{P}^{\rm I}_{te} \simeq \mathbb{P}_{fbm} + \frac{(1-\mathbb{P}_{fbm})}{2^{L_s}}  \int_0^\infty \sum_{m=0}^{L_s} \binom{L_s}{m} \Bigg\{1- \nonumber \\
&\left( 1-Q\left(  \frac{\mu  hm \left(h\mu m+2L_s\sigma_0^2\right)}
{\sigma''_{tc|h,m}}  \right) \right)^3 \Bigg\}f_h(h)dh
\end{align}
where
\begin{align}
\label{gjj3}
\sigma''^2_{tc|h,m}  =&~\left( 4L_s\sigma_0^2  + 3\mu mh  \right)^2 \left( \sigma_s^2 h  m + L_s\sigma_0^2 \right)   \\ 
&+L_s\sigma_0^2 \left( \mu mh\right)^2 
+ 2 L_s\sigma_0^2  \left(2L_s\sigma_0^2+\mu mh\right)^2. \nonumber
\end{align}
Furthermore,  eq. \eqref{TER_ICSI} can be rewritten as follows
\begin{align}
\label{TER_ICSI2}
&\mathbb{P}^{\rm I}_{te} \simeq  \mathbb{P}^{\rm I}_{te1} + \mathbb{P}^{\rm I}_{te2}
\end{align}
where 
$\mathbb{P}^{\rm I}_{te1}= \frac{7(1-\mathbb{P}_{fbm})}{2^{L_s+3}} + \mathbb{P}_{fbm}$ ,
and 
\begin{align}
&\mathbb{P}^{\rm I}_{te2} = \frac{1-\mathbb{P}_{fbm}}{2^{L_s}}  \int_0^\infty \sum_{m=1}^{L_s} \binom{L_s}{m} \Bigg\{1- \nonumber \\
&\left( 1-Q\left(  \frac{\mu  hm \left(h\mu m+2L_s\sigma_0^2\right)}
{\sigma''_{tc|h,m}}  \right) \right)^3 \Bigg\}f_h(h)dh. \nonumber
\end{align}
From \eqref{TER_ICSI2}, it can be found that unlike $\mathbb{P}^{\rm I}_{te2}$, the term $\mathbb{P}^{\rm I}_{te1}$ only depends on $L_s$ and is independent of both $h$ and the variance of the noise. Hence, at high SNR when $\mathbb{P}^{\rm I}_{te2}\ll1$, $\mathbb{P}^{\rm I}_{te}$ becomes the error floor that is equal to: maximum ${\{\frac{7}{2^{L_s+3}},\mathbb{P}_{fbm}}\}$.
%%%%%%%%%%%%%%%%%%%%%%%%%%%%%
\begin{figure}[t]
	\begin{center}
		\includegraphics[width=4 in]{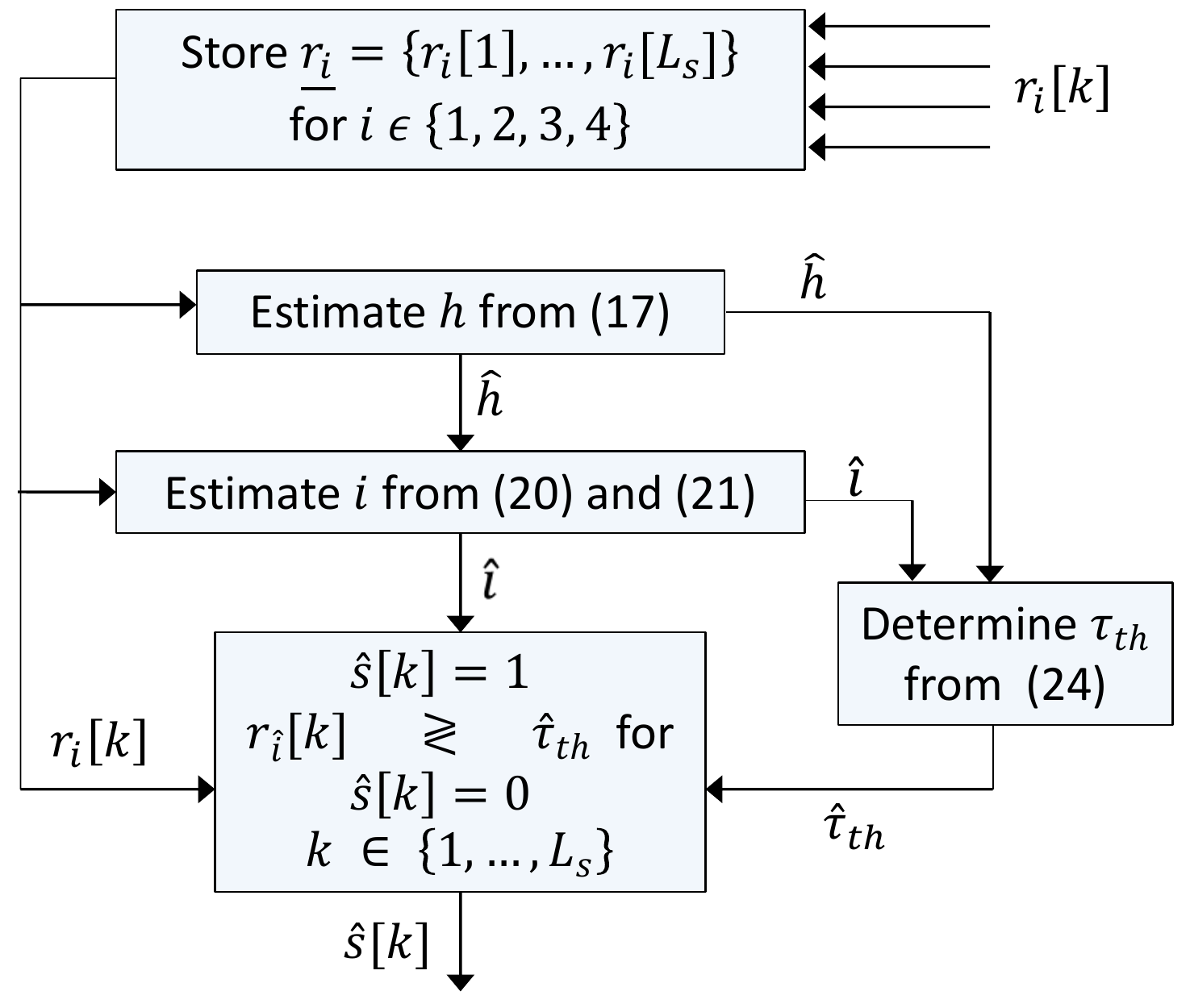}
		\caption{The flowchart of spatial beam tracking and data detection.}
		\label{flowchart}
	\end{center}
\end{figure}
%%%%%%%%%%%%%%%%%%%%%%%%%%%%
%
%%%%%%%%%%%%%%%%%%%%%%%%%%%%%%
%%%%%%%%%%%%%%%%%%%%%%%%%%%%%%
\subsection{Data Detection}
%%%%%%%%%%%%%%%%%%%%%%%%%%%%%%
%%%%%%%%%%%%%%%%%%%%%%%%%%%%%%
After performing spatial tracking, the transmitted data via OOK symbols can be detected as 
\begin{align}
\label{ds1}
r_{\hat{i}}[k]\substack{\hat{s}[k]=1 \\ >\\< \\ \hat{s}[k]=0}~ \tau_{th}(h)
\end{align}
where $\hat{i}$ is the selected APD quadrant after tracking and $\tau_{th}(h)$ denotes the detection threshold level of the OOK signaling and it can be obtained as \cite{bayaki2012edfa}
%\begin{align}
%\label{vb1}
%\tau_{th}(h) =& -\frac{  h \mu \sigma_{0}^{2}}    {\sigma_s^2 h} +
%\frac{ \sqrt{\left(\sigma_s^2 h  + \sigma_0^2\right)}\sigma_{0} }    {\sigma_s^2 h}  \nonumber\\
%& \times \sqrt{(h \mu)^{2} +2\left(\sigma_s^2 h \right)ln {\displaystyle{\left(\frac{\sigma_s^2 h  + \sigma_0^2}{\sigma_{0}^2}     \right)}}}.
%\end{align}
\begin{align}
\label{vb1}
\tau_{th}(h) =& \frac{\mu h \sigma_0}{\sigma_0+	\sqrt{h\sigma_s^2+\sigma_0^2}}.
\end{align}
From \eqref{ds1} and \eqref{vb1} it can be observed that the receiver needs to know the value of $h$ at each block of data sequence to adjust detection threshold and perform data detection afterward.
However, in practical situation, $h$ is an unknown parameter and needs to be estimated. Therefore, under unknown CSI conditions, we first estimate $\tau_{th}(h)$ by substituting \eqref{xd3} in \eqref{vb1}, and then proceed to detect transmitted data using \eqref{ds1}.

As a benchmark to evaluate our proposed data detection method, the BER of the considered system under known CSI is derived in Appendix C as
%%%%%%%%%%%%%%%%%%%%%%%%%%%%%%%%%
%%%%%%%%%%%%%%%%%%%%%%%%%%%%%%%%%
\begin{align}
\label{cx5}
\mathbb{P}^{\rm p}_{eb} &= \frac{1}{2}\mathbb{P}_{fbm} + \frac{(1-\mathbb{P}_{fbm})}{2^{L_s}}  \int_0^\infty \sum_{m=0}^{L_s} \binom{L_s}{m} \nonumber \\
&~~~\times \Bigg\{ 
\frac{m}{L_s} \left(1-Q\left(\frac{\tau_{th}(h)}{\sigma_0}\right)  \right) +\frac{L_s-m}{L_s}Q\left( \frac{\tau_{th}(h)}{\sigma_0} \right)       \nonumber \\
%%%%%%%%%%%%%%%%%%%%%%%%%%%%%%%%%%%%%%%%%%%%%%%%%%
&~~~+   \frac{m}{L_s}\left( 1-Q\left(  \frac{\mu \sigma_s^2 h^2m^2 \left(h\mu m+2L_s\sigma_0^2\right)}
{\sigma_{tc|h,m}}  \right) \right)^3 \nonumber \\	
&~~~\times \!\left(\!Q\!\left( \!\frac{\mu h\! - \!\tau_{th}(h)}{\sqrt{\sigma_s^2h+\sigma_0^2}}\right)\! +\! Q\left(\frac{\tau_{th}(h)}{\sigma_0}\right) - 1       \right) 
\Bigg\}f_h(h)dh.
\end{align} 
%%%%%%%%%%%%%%%%%%%%%%%%%%%%%%%%
%%%%%%%%%%%%%%%%%%%%%%%%%%%%%%%%
Moreover, in Appendix D, we derive the BER of the considered system under unknown CSI as
%%%%%%%%%%%%%%%%%%%%%%%%%%%%%%%%%
%%%%%%%%%%%%%%%%%%%%%%%%%%%%%%%%%
\begin{align}
\label{cn5}
\mathbb{P}^{\rm I}_{eb} &= \frac{1}{2}\mathbb{P}_{fbm} + \frac{(1-\mathbb{P}_{fbm})}{2^{L_s}}\!  \int_0^\infty \!\sum_{m=0}^{L_s} \!\binom{L_s}{m}\! \nonumber \\
&~~~\times\Bigg\{ \!
\frac{m}{L_s}\!\Bigg(\!1\!- \!Q\!\left(\!\frac{mh}{\sqrt{m\sigma_s^2h+4L_s\sigma_0^2}} \!\right) \!\Bigg)  \nonumber  \\
&~~~+\frac{L_s-m}{L_s}Q\left(\frac{mh}{\sqrt{m\sigma_s^2h+4L_s\sigma_0^2}} \right)       \nonumber \\
%%%%%%%%%%%%%%%%%%%%%%%
&~~~+ \frac{m}{L_s}\left(  1-Q\left(  \frac{\mu hm \left(h\mu m+2L_s\sigma_0^2\right)}
{\sigma''_{tc|h,m}}  \right) \right)^3 \nonumber \\
&~~~\times   
\Bigg[Q\left(\frac{C_1 L_s \mu h}{2\sqrt{C_2\left(\sigma_s^2h+\sigma_0^2\right) + C_3\left( (4L_3-1)\sigma_0^2\right)}} \right)  \nonumber \\
&~~~+ Q\left(\frac{mh}{\sqrt{m\sigma_s^2h+4L_s\sigma_0^2}} \right) - 1       \Bigg]  
\Bigg\}f_h(h)dh, 
\end{align} 
%%%%%%%%%%%%%%%%%%%%%%%%%%%%%%%%
%%%%%%%%%%%%%%%%%%%%%%%%%%%%%%%%
where
$C_1 = \frac{2m}{L_s}\sigma_s^2h +\sigma_0^2  -\left( \frac{2m-L_s}{L_s}\right)^2\sigma_0^2$,
$C_2 =  4\mu m\sigma_0^2  - \mu L_s(2\sigma_0^2  + h\sigma_s^2) - 2mL_s\left(    \sigma_s^2    +\sigma_0^2 \right)$
and $C_3 = 2\mu\sigma_0^2(2m-L_s) - \mu h\sigma_s^2L_s$.

Finally, in addition to the mathematical derivations, the flowchart for the proposed spatial beam tracking and data detection algorithms is shown in Fig. \ref{flowchart}.
%%%%%%%%%%%%%%%%%%%%%%%%%%%%%%%%%%%%%%%%%%%%%%%%%%%%%%%%%%%%%%%%

%%%%%%%%%%%%%%%%%%%%%%%%%%%%%
\begin{figure*}[t]
	\begin{center}
		\includegraphics[width=4.5 in]{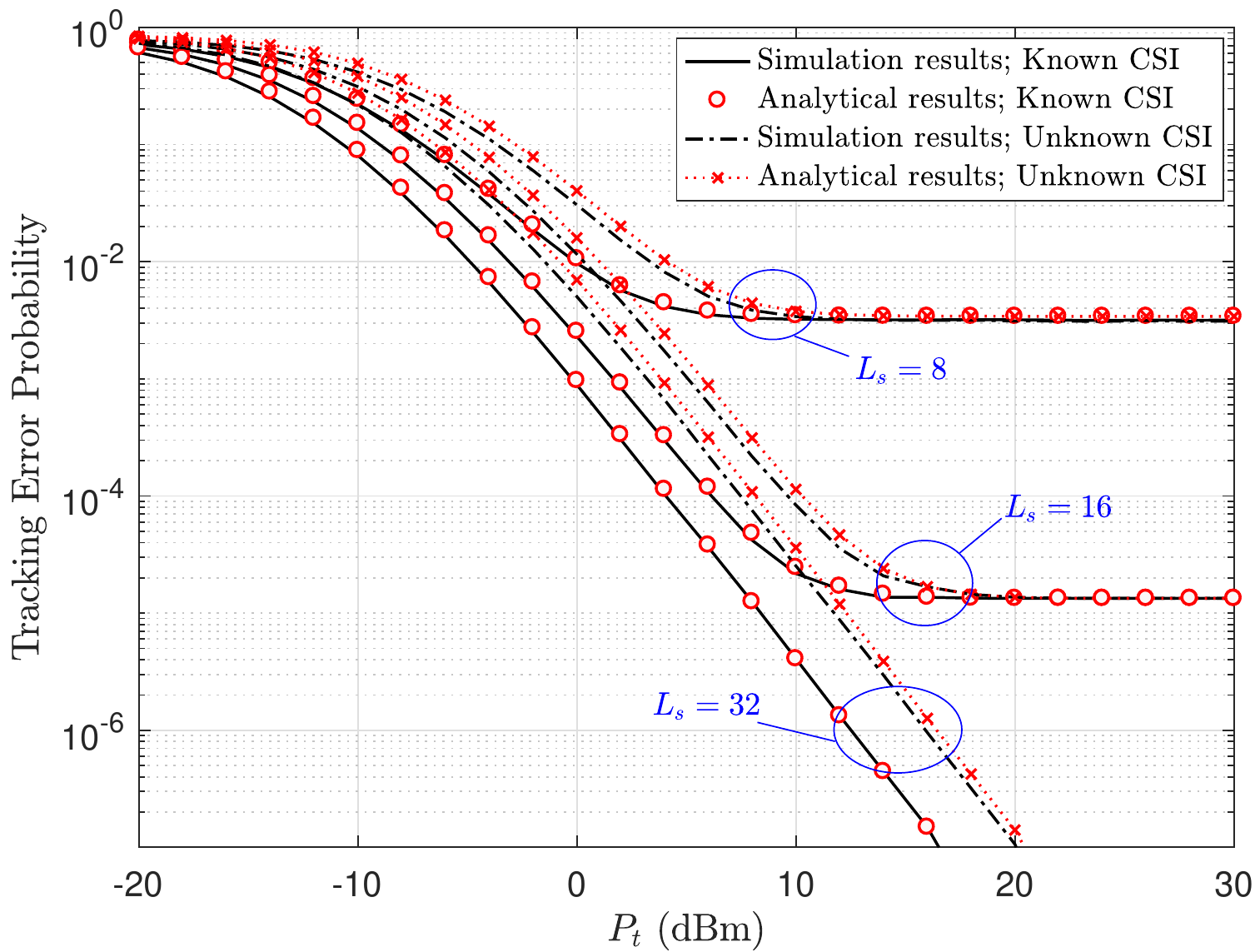}
		\caption{Tracking error probability versus $P_t$ for different values of $L_s$, $\sigma_x = \sigma_y = 4 mrad$.}
		\label{trcking-error}
	\end{center}
\end{figure*}
%%%%%%%%%%%%%%%%%%%%%%%%%%%%
%%%%%%%%%%%%%%%%%%%%%%%%%%%%%%%%%%%%%%%%%%%%%%%%%%%%%%%%%%%%%%%%
%%%%%%%%%%%%%%%%%%%%%%%%%%%%%%%%%%%%%%%%%%%%%%%%%%%%%%%%%%%%%%%%
%%%%%%%%%%%%%%%%%%%%%%%%%%%%%%%%%%%%%%%%%%%%%%%%%%%%%%%%%%%%%%%%
%%%%%%%%%%%%%%%%%%%%%%%%%%%%%%%%%%%%%%%%%%%%%%%%%%%%%%%%%%%%%%%%
%%%%%%%%%%%%%%%%%%%%%%%%%%%%%%%%%%%%%%%%%%%%%%%%%%%%%%%%%%%%%%%%
%\vspace{-1 mm}
\section{Simulation Results and Discussion}
\label{V}
%%%%%%%%%%%%%%%%%%%%%%%%%%%%%%%%%%%%%%%%%%%%%%%%%%%%%%%%%%%%%%%%
%%%%%%%%%%%%%%%%%%%%%%%%%%%%%%%%%%%%%%%%%%%%%%%%%%%%%%%%%%%%%%%%
In this section, numerical results are provided in terms of tracking error and BER to evaluate the performance of the considered methods for spatial tracking and data detection. Indeed, we carry out Monte-Carlo simulations to corroborate the accuracy of the derived analytical expressions.    Based on the practical values asserted in \cite{ghassemlooy2012optical}, the system parameters are specified in TABLE I, following our parameter definition in Section II.

We first investigate the performance of the tracking method under different length of observation window. Accordingly,  Fig. \ref{trcking-error} demonstrates tracking error probability versus $P_t$ for different values of $L_s$. Clearly, an exact match between the analytical-
and simulation-based results can be observed, which validates the accuracy of the analytical expressions in both known CSI and unknown CSI conditions. In addition, as we have anticipated, the performance of the tracking system is improved by increasing $L_s$ at the expense of more delay of tracking. Also, an error floor can be noticed in case of insufficient length of $L_s$ due to the transmission of all-zero sequences. For an observation window of length $L_s$, the occurrence probability of an all-zero sequence is equal to $\frac{1}{2^{L_s}}$. Obviously, tracking is done over noise when an all-zero sequence is transmitted. This error floor can also be realized from analytical expressions of \eqref{TER_ICSI2}.  
%On the other hand, by increasing $L_s$, the processing load and delay of the considered method and also
%the required memory linearly increase. Therefore, selecting an optimum value for $L_s$ (as we will discuss later) consists in making a
%trade-off between increasing $L_s$ to enhance system performance and decreasing $L_s$ to reduce processing time. Indeed, target BER for an FSO system is
%commonly lower than $10^{-9}$ \cite{navidpour2007ber,rockwell2001wavelength}. Obviously, at these target BERs, using Monte-Carlo simulations to determine an optimum value of $L_s$ would require a long time. Therefore, analytical expressions which are derived in this paper can be used to overcome this time-consuming problem. 
Additionally, even for large values of $L_s$, there exists a gap between the tracking methods under different scenarios of knowing CSI at the receiver. This is expected since the proposed tracking method performs in a blind way with low computational complexity.

\begin{figure*}[t]
	\begin{center}
		\includegraphics[width=6.75 in]{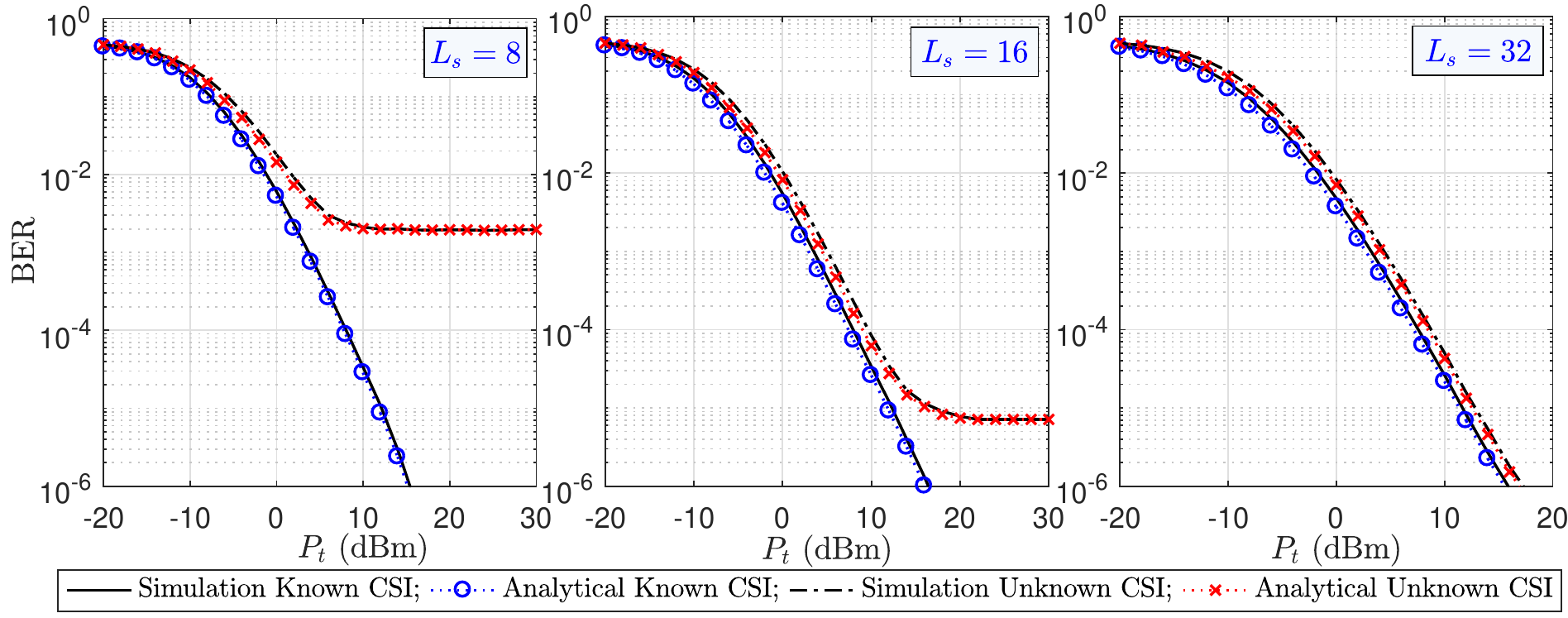}
		\caption{BER versus $P_t$ for different values of $L_s$.}
		\label{BER}
	\end{center}
\end{figure*}

We now study the BER performance of the  proposed system for tracking and data detection. Accordingly, we plot BER versus $P_t$ for different values of $L_s$ in Fig \ref{BER}.  Again, analytical calculations closely match with simulation results.  Moreover, there is no error floor when we have the knowledge of the CSI at the receiver. 
Although for all zero transmitted sequence tracking error occurs, it is not the case in the detection step. More specifically, when the receiver knows $h$, an all-zero transmitted sequence can be correctly detected since the receiver can exactly determine  $\tau_{th}$ based on \eqref{vb1}.
%However, all-zero transmitted sequence do not cause detection error.
%%%%%%%%%%
It is also clear that the performance of the system will improve by increasing the length of the observation window. When the observation window is sufficiently large, the proposed detection method under unknown CSI scenarios can achieve performance close to those achieved with known CSI. 

\begin{figure}[!]
	\begin{center}
		\includegraphics[width=4 in]{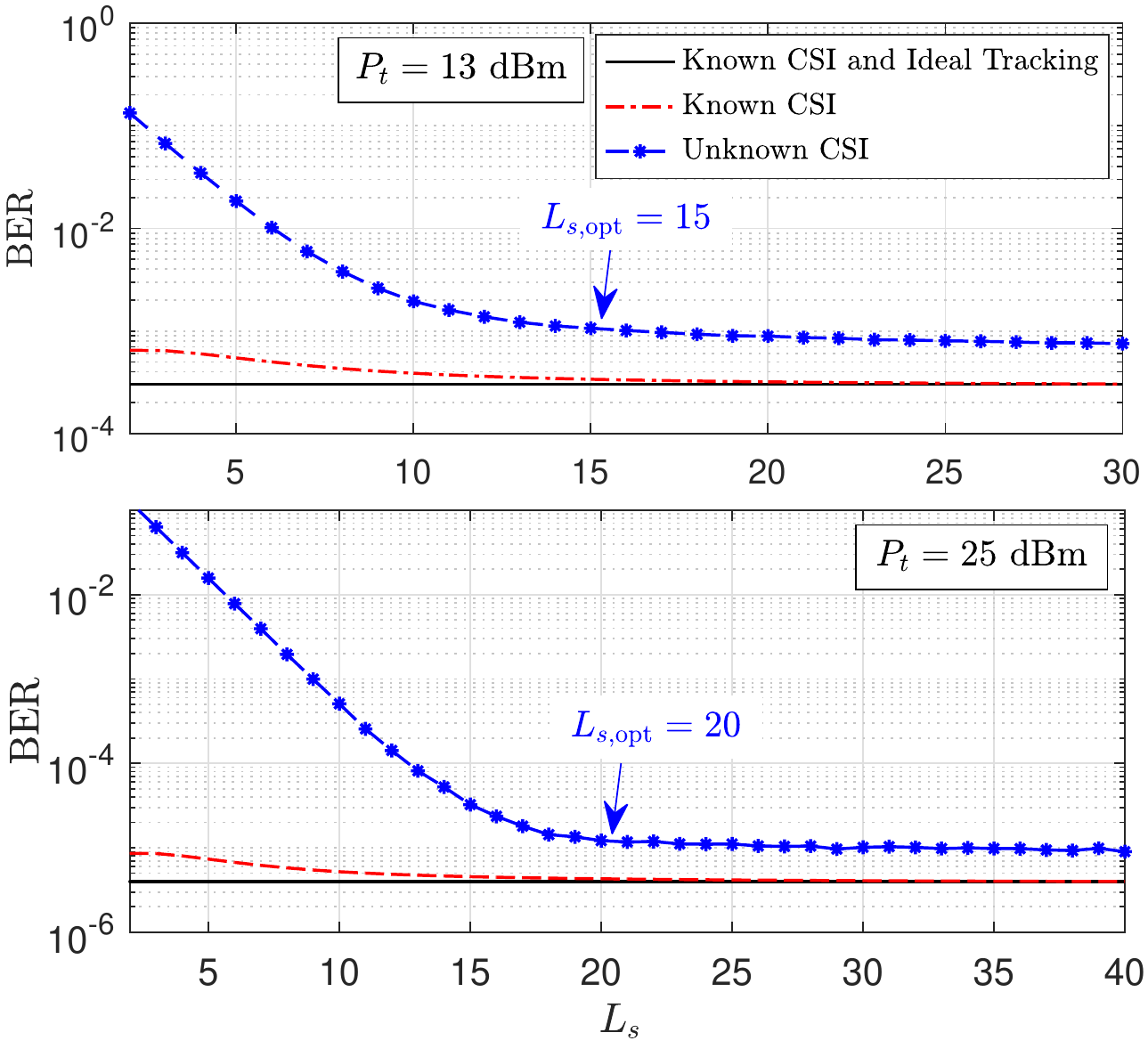}
		\caption{BER versus $P_t$ for different values of $L_s$.}
		\label{Ls-optimum}
	\end{center}
\end{figure}

To provide deeper insight into the importance of optimizing $L_s$, we have shown the BER curves of the considered methods for $P_t$ of 13 dBm and 25dBm  versus $L_s$ in Fig \ref{Ls-optimum}. Indeed, an ideal receiver with known CSI and no tracking error is considered as a lower bound benchmark. Accordingly, one can conclude that choosing the optimum value of observation window, $L_{s,opt}$, is dependent on the predetermined system parameters, i.e., tolerable delay, and desired BER. Particularly, the value of $L_s$ must be large enough to ensure that the occurrence probability of all-zero sequence is lower than the desired BER. For instance, according to Fig. \ref{Ls-optimum}, by increasing the desired BER from $10^{-3}$ to $10^{-5}$, $L_{s,opt}$ changes from 15 to 20.  It is worth mentioning that the dependence of $L_{s,opt}$ on $P_t$ implies that it is also a function of the receiver noise in practice.
%%%%%%%%%%%%%%%%%%%%%%%%%%%%%
%%%%%%%%%%%%%%%%%%%%%%%%%%%%
%%%%%%%%%%%%%%%%%%%%%%%%%%%%%%%%%%%%%%%%%%%%%%%%%%%%%%%%%%%%%%%%
%%%%%%%%%%%%%%%%%%%%%%%%%%%%%
%%%%%%%%%%%%%%%%%%%%%%%%%%%%
%%%%%%%%%%%%%%%%%%%%%%%%%%%%%%%%%%%%%%%%%%%%%%%%%%%%%%%%%%%%%%%%

\begin{figure}[!]
	\begin{center}
		\includegraphics[width=4.75 in]{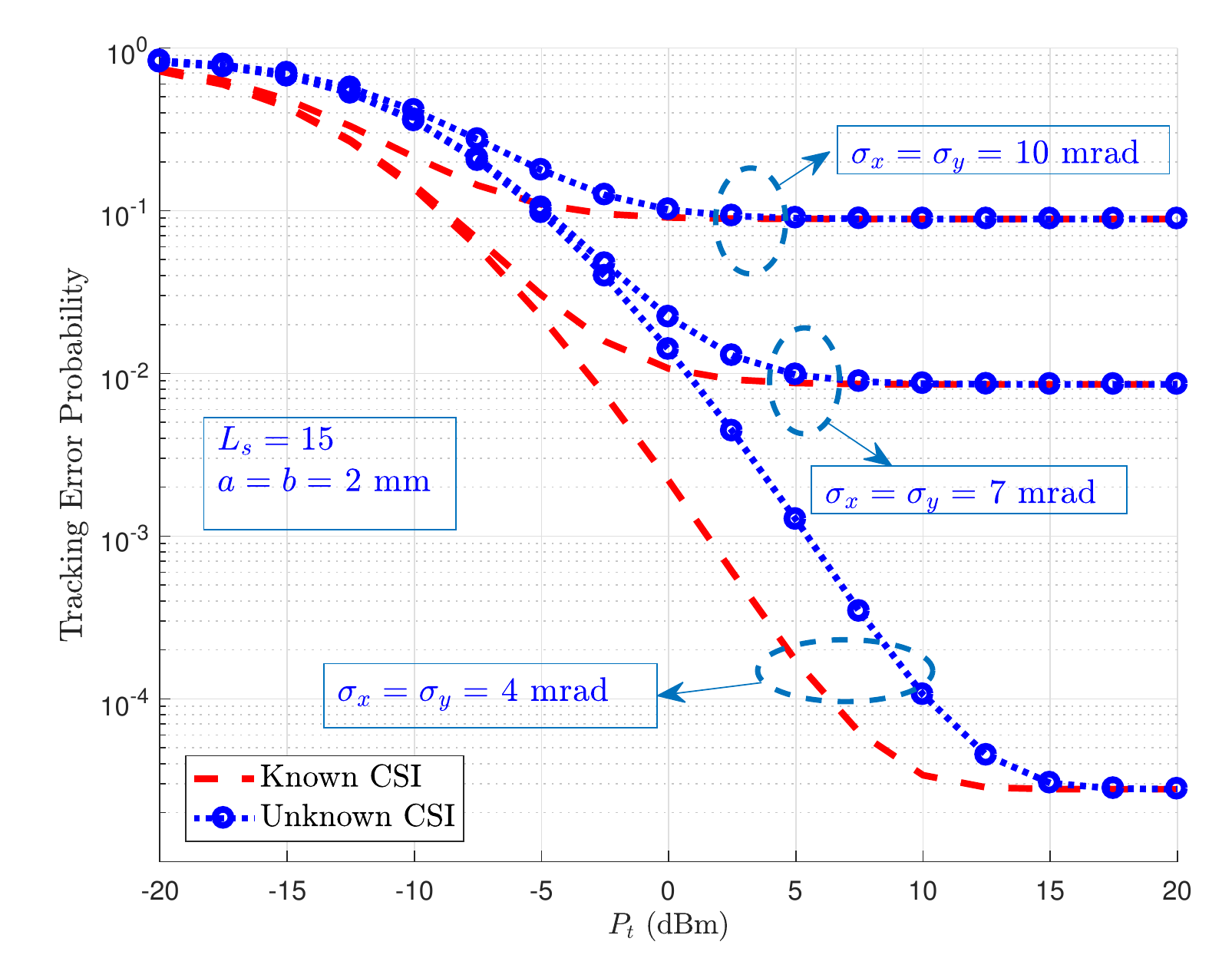}
		\caption{Tracking error probability versus $P_t$ for different values of $\sigma_x$ and $\sigma_y$.}
		\label{tracking_versus_sigma}
	\end{center}
\end{figure}

\begin{figure}[!]
	\begin{center}
		\includegraphics[width=4.75 in]{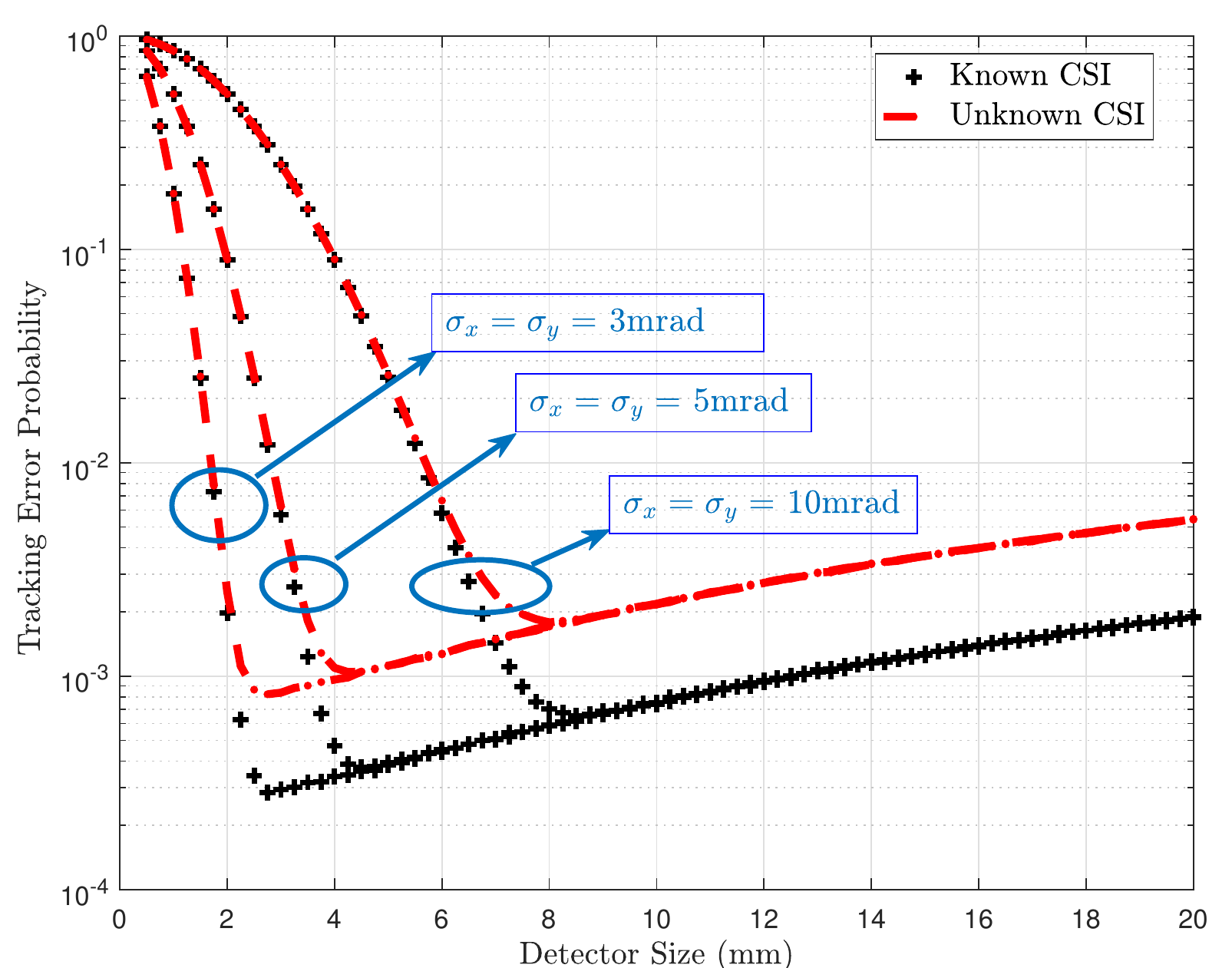}
		\caption{Tracking error probability versus size of the detector for different values of $\sigma_x$ and $\sigma_y$.}
		\label{detector_size}
	\end{center}
\end{figure}

%%%%%%%%%%%%%%%%%%%%%%%%%%%%%%%%%%%%%%%%%%%%%%%%%%%%%%%%%%%%%%%%
\begin{table}
	\label{sys-parameters}
	\caption{System Parameters Used Throughout Simulations} % title of Table
	\centering % used for centering table
	\begin{tabular}{c c c} % centered columns (3 columns)
		\hline\hline \\[-.5ex]%inserts double horizontal lines\\
		Name & Parameter & Value \\ [.5ex] % inserts table
		%heading
		\hline\hline \\[-1.2ex]% inserts single horizontal line
		APD Gain                        & $ G $           &$ 100 $  \\[1ex] 
		Quantum Efficiency              & $ \eta $         & $ 0.9 $ \\[1ex]
		Avalanche Unization Factor      & $ k_{eff} $     & $ 0.028 $ \\[1ex]
		Plank\textquotesingle s Constant&$ h_{p} $        &$ 6.6 \times 10^{-34} $  \\[1ex]
		Wavelength                      &$ \lambda $      &$ 1550$ nm   \\[1ex]               
		Optical Frequency               & $ \nu $         & $ 1.93 \times 10^{14} $ \\[1ex]              
		Boltzmann\textquotesingle s Constant&$ K_{B} $    & $ 1.38 \times 10^{-22} $ \\[1ex]                
		Receiver Load                   & $ R_{l} $       & $ 1~k\Omega $ \\[1ex]                
		Receiver Temperature            &$ T_{r} $        & $ 300\degree~K$  \\[1ex]                
		Bit Time                        &$ T_{b} $        & $ 10^{-9} $ \\[1ex]  
%		Extinction Ratio                &$ \alpha_e $     & $ 0.2 $ \\[1ex]               
		Aperture Radius                 & $ r $           & $ 5$ cm \\[1ex] 
		Normalized Beam Width           & $ w_L/r $       & $ 12 $\\[1ex] 
		Normalized Jitter               & $ \sigma_j/r $  & $ 2 $\\[1ex]             
		Background Power                & $ P_{b} $       & $ 100$ nW  \\[1ex]                
%		Channel Length                  & $ d_{0} $       &$ 1 $ km   \\[1ex]                               
		Rytov Variance                  &$ \chi^{2} $ & $ 1$ \\[.1ex]
		\hline\hline %inserts single line
	\end{tabular}
	\end{table}
%%%%%%%%%%%%%%%%%%%%%%%%%%%%%%%%%%%%%%%%%%%%%%%%%%
%%%%%%%%%%%%%%%%%%%%%%%%%%%%%%%%%%%%%%%%%%%%%%%%%% 
To have a deeper understanding about the effect of  hovering fluctuations of the receiver on the link performance, we have depicted tracking error probability versus $P_t$ for different values of $\sigma_x$ and $\sigma_y$ in Fig. \ref{tracking_versus_sigma}. Note that, the degree of instabilities of the hovering UAV is considered in the order of several \rm{mrad} owing to the invention of mechanical and control systems for UAVs with high accuracy \cite{orsag2017dexterous}. As expected, tracking error probability increases via increasing AoA fluctuations at the receiver side.
%%%%%%%%%%%%%%%%%%%%%%%%%%%%%
%%%%%%%%%%%%%%%%%%%%%%%%%%%%
However, such performance degradation can be improved  by increasing $\Phi_{FoV}$ via enlarging the size of photo detector and by avoiding full beam misalignment. On the other hand, to reduce the effect of undesired background noise, the area of photo detector should be as small as possible. Regarding this trade-off,  tracking error probability versus size of the detector for different values of $\sigma_x$ and $\sigma_y$ is depicted in Fig. \ref{detector_size}. Without loss of generality, in this figure we assume that the sides of detector are equal, i.e., $a = b$. As we can observe from Fig. \ref{detector_size}, choosing an optimum size for the detector can considerably alleviate the impacts of hovering fluctuations on the performance of the tracking method. However, the electrical bandwidth of a photo detector will decrease by enlarging its size, and meanwhile the amount of undesired background noise due to a larger FoV can adversely affect the system performance. 
%%%%%%%%%%%%%%%%%%%%%%%%%%%%%
%%%%%%%%%%%%%%%%%%%%%%%%%%%%
%we assume a Ground to UAV link in which optical transmitter is mounted on a ground station, hence its orientation deviations can be ignored. However, at the receiver side, we results in have AoA fluctuations 

%%%%%%%%%%%%%%%%%%%%%%%%%%%%%%%%%%%%%%%%%%%%%%%%%%%%%%%%%%%%%%
%%%%%%%%%%%%%%%%%%%%%%%%%%%%%%%%%%%%%%%%%%%%%%%%%%%%%%%%%%%%%%
%\vspace{-2 mm}
\section{Conclusion}
%%%%%%%%%%%%%%%%%%%%%%%%%%%%%%%%%%%%%%%%%%%%%%%%%%%%%%%%%%%%%%
%%%%%%%%%%%%%%%%%%%%%%%%%%%%%%%%%%%%%%%%%%%%%%%%%%%%%%%%%%%%%%
In this paper, for an FSO link to a hovering UAV, we assumed a practical scenario in which the receiver does not know CSI, and then investigated beam tracking and data detection for the case of OOK signaling, in the presence of random hovering fluctuations of the UAV. For optimal OOK signal demodulation, the receiver requires the knowledge of the instantaneous CSI. Therefore, incorporating sequential received OOK symbols, we first determined the direction of arrival of the received optical beam at the receiver and then estimated CSI blindly to increase the link bandwidth efficiency.  Consequently, data detection was performed using the estimated channel coefficient. We also provided detailed mathematical analysis and derived closed-form formulations of tracking error and BER to evaluate the performance of the considered system. It was shown that the hovering fluctuations have dramatic impact on beam tracking. However, choosing an optimum size of PDs can alleviate such performance degradation. Also, the optimized value of $L_s$ achieves a compromise between desired performance of tracking along with data detection methods and tolerable complexity of the system.  The high accuracy of the analytical analysis was verified by using Monte-Carlo simulations. Our results can thus be used to determine the optimum value of the detector size as well as the length of observation window $L_s$ without resorting to laborious Monte-Carlo simulation. 
%we considered a quad-detector arrangement of APDs at the receiver for optical beam tracking and investigated the effect of random hovering fluctuations of the UAV on the performance of the tracking method. Particularly, 

%which enables system designers to calcu and tuning of optimal $L_s$ and the detector size in an adaptive system, thus removing the need to complex and time consuming Monte
%Carlo simulations.
%%%%%%%%%%%%%%%%%%%%%%%%%%%%%%%%%%%%%%%%%%%%%%%%%%%%%%%%%%%%%%
\appendices

%%%%%%%%%%%%%%%%%%%%%%%%%%%%%%%%%%%%%%%%%%%%%%%%%%%%%%%%%%%%%%
\section{Tracking Error Analysis Under Known CSI}
%%%%%%%%%%%%%%%%%%%%%%%%%%%%%%%%%%%%%%%%%%%%%%%%%%%%%%%%%%%%%%%
Tracking error of the considered system is expressed as 
{
\begin{align}
\label{kd111}
\mathbb{P}_{te} = \mathbb{P}_{fbm} + (1-\mathbb{P}_{fbm})\int_0^\infty \mathbb{P}_{te|h} f_h(h) dh
\end{align}
}
where 
\begin{align}
\label{kd222}
\mathbb{P}_{te|h} = \sum_{m=0}^{L_s} p(m) \mathbb{P}^{\rm p}_{te|h,m}
\end{align}
and 
\begin{align}
\label{kd3}
\mathbb{P}^{\rm p}_{te|h,m} = 1-\mathbb{P}_{tc|h,m}^{\rm p}.
\end{align}
In \eqref{kd222} and \eqref{kd3}, $\mathbb{P}^{\rm p}_{te|h,m}$ and $\mathbb{P}^{\rm p}_{tc|h,m}$, respectively, denote the tracking error probability  and the probability of correct tracking conditioned on $h$ and $m$. Also, $ P(m) = \binom{L_{s}}{m}/2^{L_s} $ is the probability that $ m $ bits out of $ L_s $ transmitted bits are equal to one. To calculate $\mathbb{P}_{tc|h,m}$, without loss of generality, we assume that  the first quadrant is the target PD, i.e., $D_1 = 1$. Hence, $\mathbb{P}_{tc|h,m}$ is the probability that $\mathcal{M}_{1|h,m}$ is lower than $\mathcal{M}_{j|h,m}$ for $j = 2,3,4$. Accordingly, $\mathbb{P}_{tc|h,m}$ can be obtained as
\begin{align}
\label{kd4}
\mathbb{P}_{tc|h,m}^{\rm p} = {\rm Prob}\left\{\mathcal{M}_{1|h,m} < \mathcal{M}_{\min|h,m} \right\}
\end{align}
where 
\begin{align}
\label{kdd1}
\mathcal{M}_{\min|h,m} = \min\left(\mathcal{M}_{2|h,m},\mathcal{M}_{3|h,m},\mathcal{M}_{4|h,m}\right).
\end{align}
Since the noises of the APDs are independent, eq. \eqref{kd4} can be obtained as
\begin{align}
\label{kd5}
\mathbb{P}_{tc|h,m}^{\rm p} = \left(\mathbb{P}'^{\rm p}_{tc|h,m}\right)^3
\end{align}
where
\begin{align}
\label{kd6}
\mathbb{P}'^{\rm p}_{tc|h,m} &= {\rm Prob}\left\{\mathcal{M}_{1|h,m} < \mathcal{M}_{2|h,m} \right\}  \\
&= {\rm Prob}\left\{\mathcal{M}_{1|h,m} < \mathcal{M}_{3|h,m} \right\} \nonumber \\
&= {\rm Prob}\left\{\mathcal{M}_{1|h,m} < \mathcal{M}_{4|h,m} \right\}.\nonumber
\end{align}
Substituting \eqref{metric-tracking-2} into \eqref{kd6}, $\mathbb{P}'^{\rm p}_{tc|h,m}$ can be obtained as \eqref{kd7}.
%
%%%%%%%%%%%%%%%%%%%%%%%%%%%%%%%%%
%%%%%%%%%%%%%%%%%%%%%%%%%%%%%%%%%
\begin{figure*}[b]
	\normalsize
	\hrulefill
	\begin{align}
	\label{kd7}
	\mathbb{P}'^{\rm p}_{tc|h,m} &= {\rm Prob}\left\{ 
	\frac{{{{\left| {{{r'}_{1|h,m}} - h\mu m} \right|}^2}}}{{\sigma _s^2hm + {L_s}\sigma _0^2}}
	+\sum_{j=2}^4\!{\frac{{{{\left| {{{r'}_{j|h,m}}} \right|}^2}}}{{{L_s}\sigma _0^2}}} <
	%%%%%
	\frac{{{{\left| {{{r'}_{2|h,m}} - h\mu m} \right|}^2}}}{{\sigma _s^2hm + {L_s}\sigma _0^2}} +
	\frac{{{{\left| {{{r'}_{1|h,m}}} \right|}^2}}}{{{L_s}\sigma _0^2}}
	+\sum_{j=3,4}{\frac{{{{\left| {{{r'}_{j|h,m}}} \right|}^2}}}{{{L_s}\sigma _0^2}}} 
	\right\} \nonumber \\
	%%%%%%%%%%%%%%%%%%%%%%%%%%%%%%%%%%%%%%%%%%
	&= {\rm Prob}\left\{ 
	\sigma_s^2hm \left(  \left( r'_{1|h,m}  \right)^2  -  \left( r'_{2|h,m}  \right)^2 \right)
	+ 2mL_s\sigma_0^2 \sigma_s^2 h \left(   r'_{1|h,m}  -  r'_{2|h,m}  \right) >0
	\right\}.
	\end{align} 
	%\hrulefill
\end{figure*}
%%%%%%%%%%%%%%%%%%%%%%%%%%%%%%%%
%%%%%%%%%%%%%%%%%%%%%%%%%%%%%%%%
%$r'_{1|h,m,{D_1} = 1} - h\mu m$
%$\sigma_s^2hm + L_s\sigma_0^2$
From \eqref{conditioned-r}, we rewrite \eqref{kd7} as
\begin{align}
\label{kd12}
&\mathbb{P}'^{\rm p}_{tc|h,m}  \nonumber \\ 
&=\!{\rm Prob}\bigg\{\!  
\sigma_s^2hm \!\left(  \left( h\mu m + n'_{1|h,m,D_1=1}\right)^2 \right. 
\!\!-\! \! \left. \left( n'_{2|h,m,D_2=0}  \right)^2 \right) \nonumber \\
%%%
&~~+ \!2mL_s\sigma_0^2 \sigma_s^2 h \left( h\mu m + n'_{1|h,m,D_1=1}- n'_{2|h,m,D_2=0} \right) >0
\bigg\} \nonumber \\
%%%%%%%%%%%%%%%%%%%%%%%%%%%%%%%%%%%%%%%%%%%
&=\!{\rm Prob}\bigg\{\!
\mu \sigma_s^2 h^2m^2\!\left(h\mu m+2L_s\sigma_0^2\right)\! + \!n'_{tc|h,m} >0\bigg\},
\end{align}
where
\begin{align}
\label{hj1}
n'_{tc|h,m} =&~ \sigma_s^2hm \left(  \left( n'_{1|h,m,D_1=1}\right)^2  
- \left( n'_{2|h,m,D_2=0}  \right)^2 \right) \nonumber \\
&+2\sigma_s^2 h m \left(h\mu m + L_s\sigma_0^2 \right) n'_{1|h,m,D_1=1} \nonumber \\
&-2m L_s\sigma_0^2\sigma_s^2 h  n'_{2|h,m,D_2=0}.
\end{align}                  
At high SNR,  we have $\left( n'_{1|h,m,D_1=1}\right)^2\ll n'_{1|h,m,D_1=1}$ and 
$\left( n'_{2|h,m,D_2=0}\right)^2\ll n'_{2|h,m,D_2=0}$. Hence,  eq. \eqref{hj1} can be approximated as
\begin{align}
\label{hj2}
n'_{tc|h,m} \simeq&~ 2\sigma_s^2 h m \left(h\mu m + L_s\sigma_0^2 \right) n'_{1|h,m,D_1=1} \nonumber \\
&-2m L_s\sigma_0^2\sigma_s^2 h  n'_{2|h,m,D_2=0}.
\end{align} 					
According to \eqref{hj2} and \eqref{variance-noise-i}, $n'_{tc|h,m}$ is approximately Gaussian distributed with zero mean and variance 									
\begin{align}
\label{hj3}
\sigma_{tc|h,m}^2  =&~\left(2\sigma_s^2 h m \left(h\mu m + L_s\sigma_0^2 \right)\right)^2 
\times \left(\sigma_s^2 h m + L_s\sigma_0^2\right)  \nonumber \\ 
&~+L_s\sigma_0^2   \left(2mL_s\sigma_0^2 \sigma_s^2 h\right)^2  .
\end{align}
Based on \eqref{hj3} and \eqref{kd12}, $\mathbb{P}'^{\rm p}_{tc|h,m}$ can be derived as
\begin{align}
\label{hj4}
\mathbb{P}'^{\rm p}_{tc|h,m} \simeq 1-Q\left(  \frac{\mu \sigma_s^2 h^2m^2 \left(h\mu m+2L_s\sigma_0^2\right)}
{\sigma_{tc|h,m}}  \right).
\end{align}
Finally, by substituting \eqref{hj4}, \eqref{kd5}, \eqref{kd3}, and \eqref{kd222} into \eqref{kd111}, under known CSI, the closed form expression of  tracking error probability is obtained in \eqref{TER_PCSI}.

%%%%%%%%%%%%%%%%%%%%%%%%%%%%%%%%%%%%%%%%%%%%%%%%%%%%%%%%%%%%%%
\section{Tracking Error Analysis Under Unknown CSI}
%%%%%%%%%%%%%%%%%%%%%%%%%%%%%%%%%%%%%%%%%%%%%%%%%%%%%%%%%%%%%%%
For \eqref{kd3}, we need to calculate the correct tracking probability under unknown CSI.
According to \eqref{xd6}, \eqref{xd7} and similar to the derivation of \eqref{kd5}, the correct tracking probability conditioned on $h$ and $m$ is obtained as
\begin{align}
\label{fd5}
\mathbb{P}_{tc|h,m}^{\rm I} = \left(\mathbb{P}'^{\rm I}_{tc|h,m}\right)^3
\end{align}
where
\begin{align}
\label{fd6}
\mathbb{P}'^{\rm I}_{tc|h,m} &= {\rm Prob}\left\{\mathcal{M}'_{1|h,m} < \mathcal{M}'_{2|h,m} \right\}.
\end{align}
Substituting \eqref{xd7} into \eqref{fd6} and after some manipulations, we have
%%%%%%%%%%%%%%%%%%%%%%%%%%%%%%%%%
%%%%%%%%%%%%%%%%%%%%%%%%%%%%%%%%%
\begin{align}
\label{fd7}
\mathbb{P}'^{\rm I}_{tc|h,m} \!&= \!{\rm Prob}\Bigg\{\! 
\sigma_s^2\hat{h}m \left(  \left( r'_{1|h,m,{D_1} = 1}  \right)^2 \! -\!  \left( r'_{2|h,m,{D_2} = 0}  \right)^2 \right) \nonumber \\
%%%%%%
&+ 2mL_s\sigma_0^2 \sigma_s^2 \hat{h} \left(   r'_{1|h,m,{D_1} = 1}  -  r'_{2|h,m,{D_2} = 0}  \right) >0
\!\Bigg\}.
\end{align} 
%%%%%%%%%%%%%%%%%%%%%%%%%%%%%%%%
%%%%%%%%%%%%%%%%%%%%%%%%%%%%%%%%
Substituting \eqref{conditioned-r} and \eqref{xd3} into \eqref{fd7} and ignoring the effect of second- and third-order noise (which is a reasonable assumption at high SNR), $\mathbb{P}'^{\rm I}_{tc|h,m}$ can be approximated as
\begin{align}
\label{fd12}
&\mathbb{P}'^{\rm I}_{tc|h,m}   
%%%%%%%%%%%%%%%%%%%%%%%%%%%%%%%%%%%%%%%%%%%
\simeq{\rm Prob}\big\{
\mu m h(\mu m h + 2L_s\sigma_0^2 ) + n''_{tc|h,m} >0\big\}
\end{align}  
where
\begin{align}
\label{fd14}
n''_{tc|h,m} =& \!\left( 4L_s\sigma_0^2  \!+ \!3\mu mh  \right)\!n'_{1|h,m,D_1=1}   
+ \mu mhn'_{2|h,m,D_2=0} \nonumber \\
&+  \left(2L_s\sigma_0^2+\mu mh\right) \sum_{i=3}^{4}n'_{i|h,m,D_i=0}.
\end{align}
From \eqref{hj2} and \eqref{variance-noise-i}, variance of $n''_{tc|h,m}$ can be approximated as	
\begin{align}
\label{gj3}
\sigma''^2_{tc|h,m}  =&~\left( 4L_s\sigma_0^2  + 3\mu mh  \right)^2 \left( \sigma_s^2 h  m + L_s\sigma_0^2 \right)   \\ 
&+L_s\sigma_0^2 \left( \mu mh\right)^2 
+ 2 L_s\sigma_0^2  \left(2L_s\sigma_0^2+\mu mh\right)^2.\nonumber
\end{align}
Based on \eqref{gj3} and \eqref{fd12}, $\mathbb{P}'^{\rm I}_{tc|h,m}$ is derived as
\begin{align}
\label{gj4}
\mathbb{P}'^{\rm I}_{tc|h,m} = 1-Q\left(  \frac{\mu hm \left(h\mu m+2L_s\sigma_0^2\right)}
{\sigma''_{tc|h,m}}  \right).
\end{align}
Finally, by substituting \eqref{gj4}, \eqref{fd5}, \eqref{kd3} and \eqref{kd2} into \eqref{kd111}, the closed-form expression of tracking error probability  for the proposed method under unknown CSI is attained in \eqref{TER_ICSI}.

%%%%%%%%%%%%%%%%%%%%%%%%%%%%%%%%%%%%%%%%%%%%%%%%%%%%%%%%%%%%%%
\section{BER Analysis Under Known CSI}
%%%%%%%%%%%%%%%%%%%%%%%%%%%%%%%%%%%%%%%%%%%%%%%%%%%%%%%%%%%%%%%
We have
{
	\begin{align}
	\label{kd1}
	\mathbb{P}_{eb} = \frac{1}{2}\mathbb{P}_{fbm} + (1-\mathbb{P}_{fbm})\int_0^\infty \mathbb{P}_{eb|h} f_h(h) dh
	\end{align}
where 
\begin{align}
\label{kd2}
\mathbb{P}_{eb|h} = \sum_{m=0}^{L_s} p(m) \mathbb{P}^{\rm p}_{eb|h,m}
\end{align}
}
Given $h$ and $m$, the BER of the considered system depends on the tracking error probability and it can be written as
\begin{align}
\label{cx1}
\mathbb{P}^{\rm p}_{eb|h,m} = \mathbb{P}^{\rm p}_{te|h,m}\mathbb{P}^{\rm p}_{e|h,m,e}+\mathbb{P}^{\rm p}_{tc|h,m}\mathbb{P}^{\rm p}_{e|h,m,c} ~
\end{align}
where $\mathbb{P}^{\rm p}_{e|h,m,e}$ is the BER conditioned on $h$ and $m$ when the tracking error occurred and  $\mathbb{P}^{\rm p}_{e|h,m,c}$ is the BER conditioned on $h$ and $m$ when the photo detector is correctly selected (no tracking error).
From \eqref{kd3}, eq. \eqref{cx1} can be rewritten as
\begin{align}
\label{cxx1}
\mathbb{P}^{\rm p}_{eb|h,m} =\mathbb{P}^{\rm p}_{e|h,m,e}
+\mathbb{P}^{\rm p}_{tc|h,m}\left(\mathbb{P}^{\rm p}_{e|h,m,c}-\mathbb{P}^{\rm p}_{e|h,m,e}\right).
\end{align}
When tracking error occurs, the receiver decides based on the noise, and hence we have
\begin{align}
\label{cx2}
&\mathbb{P}^{\rm p}_{e|h,m,e} = {\rm Prob}\left\{ n_i[k]<\tau_{th}(h), s[k]=1\right\} \nonumber \\
&~~~+{\rm Prob}\left\{ n_i[k]>\tau_{th}(h), s[k]=0\right\} \nonumber \\
%%%%%%%%
&= p(s[k]\!=\!1)\times{\rm Prob}\left\{ n_i[k]<\tau_{th}(h)|~ s[k]=1\right\}  \nonumber \\
&~~~+p(s[k]\!=\!0)\times {\rm Prob}\left\{ n_i[k]>\tau_{th}(h)| s[k]=0\right\} \nonumber \\
%%%%%%%%
&= \frac{m}{L_s}\times{\rm Prob}\left\{ n_i[k]<\tau_{th}(h)~| s[k]=1\right\} \nonumber \\
&~~~+\frac{L_s-m}{L_s}\times {\rm Prob}\left\{ n_i[k]>\tau_{th}(h)~| s[k]=0\right\} \nonumber \\
%%%%%%%%
&=\frac{m}{L_s} \left(1-Q\left(\frac{\tau_{th}(h)}{\sigma_0}\right)  \right)
+\frac{L_s-m}{L_s}Q\left( \frac{\tau_{th}(h)}{\sigma_0} \right). 
\end{align} 
When the photo detector is correctly selected, we have
\begin{align}
\label{cx3}
\mathbb{P}^{\rm p}_{e|h,m,c} &= \!\frac{m}{L_s}\times{\rm Prob}\!\left\{\! \mu h + n_i[k]<\tau_{th}(h)\big| s[k]=1\!\right\} \\
&~~~+\frac{L_s-m}{L_s}\times {\rm Prob}\left\{n_i[k]>\tau_{th}(h)\big| s[k]=0\right\} \nonumber \\
%%%%%%%%
&=\!\frac{m}{L_s} \left(\!Q\left(\! \frac{\mu h\! -\! \tau_{th}(h)}{\sqrt{\sigma_s^2h\!+\!\sigma_0^2}}\!\right)\!  \right) \!
+\!\frac{L_s\!-\!m}{L_s}\!Q\left(\! \frac{\tau_{th}(h)}{\sigma_0}\! \right)\!. \nonumber
\end{align}
Now, by substituting \eqref{cx2}, \eqref{cx3} and \eqref{kd5} into \eqref{cxx1} and by using \eqref{kd2} and \eqref{kd1}, the BER of the considered system with the perfect knowledge of $h$ is derived in \eqref{cx5}.
%%%%%%%%%%%%%%%%%%%%%%%%%%%%%%%%%%%%%%%%%%%%%%%%%%%
%%%%%%%%%%%%%%%%%%%%%%%%%%%%%%%%%%%%%%%%%%%%%%%%%%%
%%%%%%%%%%%%%%%%%%%%%%%%%%%%%%%%%%%%%%%%%%%%%%%%%%%%%%%%%%%%%%
\section{BER Analysis Under Unknown CSI}
%%%%%%%%%%%%%%%%%%%%%%%%%%%%%%%%%%%%%%%%%%%%%%%%%%%%%%%%%%%%%%%
Under unknown CSI, BER conditioned on $h$ and $m$ can be obtained as
\begin{align}
\label{cv1}
\mathbb{P}^{\rm I}_{eb|h,m} = \mathbb{P}^{\rm I}_{te|h,m}\mathbb{P}^{\rm I}_{e|h,m,e}+\mathbb{P}^{\rm I}_{tc|h,m}\mathbb{P}^{\rm I}_{e|h,m,c} ~
\end{align}
where $\mathbb{P}^{\rm I}_{e|h,m,e}$ is the BER conditioned on $h$ and $m$ when the tracking error occurs and  $\mathbb{P}^{\rm I}_{e|h,m,c}$ is the BER conditioned on $h$ and $m$ when the photo detector is correctly selected.
From \eqref{kd3}, eq. \eqref{cv1} can be rewritten as
\begin{align}
\label{cv2}
\mathbb{P}^{\rm I}_{eb|h,m} =\mathbb{P}^{\rm I}_{e|h,m,e}
+\mathbb{P}^{\rm I}_{tc|h,m}\left(\mathbb{P}^{\rm I}_{e|h,m,c}-\mathbb{P}^{\rm I}_{e|h,m,e}\right).
\end{align}
When the tracking is performed incorrectly, the decision is made based on the noise at the receiver, and hence we have
\begin{align}
\label{cv3}
\mathbb{P}^{\rm I}_{e|h,m,e} &= \frac{m}{L_s}{\rm Prob}\left\{ n_1[k]<\tau_{th}(\hat{h})~| s[k]=1\right\} \nonumber \\
&~~+\frac{L_s-m}{L_s}{\rm Prob}\left\{ n_1[k]>\tau_{th}(\hat{h})~| s[k]=0\right\} \nonumber \\
%%%%%%%%%%%%%%%%%%%%%%%%
&= \frac{m}{L_s}{\rm Prob}\left\{ n_1[k]<\frac{\mu\hat{h} \sigma_0}{\sigma_0 + \sqrt{\hat{h}\sigma_s^2+\sigma_0^2}}~\Bigg| s[k]=1\right\} \nonumber \\
&~~+\!\frac{L_s\!-\!m}{L_s}\!{\rm Prob}\!\left\{ \!n_1[k]>\!\frac{\mu\hat{h} \sigma_0}{\sigma_0 \!+\! \sqrt{\hat{h}\sigma_s^2\!+\!\sigma_0^2}}\Bigg|\! s[k]\!=\!0\!\right\}\!. 
\end{align} 
Moreover, with the assumption of high SNR or equivalently at low values of noise, $\mathbb{P}^{\rm I}_{e|h,m,e} $ can be approximated as
\begin{align}
\mathbb{P}^{\rm I}_{e|h,m,e} &\simeq  \frac{m}{L_s}\left(1-Q\left(\frac{mh}{\sqrt{m\sigma_s^2h+4L_s\sigma_0^2}} \right) \right) \nonumber \\
%%%%%%%%%%%%%%%%%%%
&~~+\frac{L_s-m}{L_s}Q\left(\frac{mh}{\sqrt{m\sigma_s^2h+4L_s\sigma_0^2}} \right). 
\end{align}  
Accordingly, from \eqref{ds1}, \eqref{vb1} and \eqref{xd2} and after some simplifications, at high SNR, $\mathbb{P}^{\rm p}_{e|h,m,c}$ can be obtained as
\begin{align}
\label{ccx3}
\mathbb{P}^{\rm p}_{e|h,m,c} &\simeq \frac{m}{L_s}{\rm Prob}\left\{ r_1[k]<\frac{\mu\hat{h} \sigma_0}{\sigma_0 + \sqrt{\hat{h}\sigma_s^2+\sigma_0^2}}~\Bigg| s[k]=1\right\} \nonumber \\
&~~+\!\frac{L_s\!-\!m}{L_s}\!{\rm Prob}\!\left\{ \!n_1[k]>\!\frac{\mu\hat{h} \sigma_0}{\sigma_0 \!+\! \sqrt{\hat{h}\sigma_s^2\!+\!\sigma_0^2}}\Bigg|\! s[k]\!=\!0\!\right\} \nonumber \\
%%%%
&=\frac{m}{L_s}Q\left(\frac{C_1 L_s \mu h}{2\sqrt{C_2\left(\sigma_s^2h+\sigma_0^2\right) + C_3\left( (4L_3-1)\sigma_0^2\right)}} \right)  \nonumber \\
%%%%%%%%%%%%%%%%%%%
&~~+\frac{L_s-m}{L_s}Q\left(\frac{mh}{\sqrt{m\sigma_s^2h+4L_s\sigma_0^2}} \right)
\end{align}
where
$C_1 = \frac{2m}{L_s}\sigma_s^2h +\sigma_0^2  -\left( \frac{2m-L_s}{L_s}\right)^2\sigma_0^2$,
$C_2 =  4\mu m\sigma_0^2  - \mu L_s(2\sigma_0^2  + h\sigma_s^2) - 2mL_s\left(    \sigma_s^2    +\sigma_0^2 \right)$
and $C_3 = 2\mu\sigma_0^2(2m-L_s) - \mu h\sigma_s^2L_s$.
Finally, by substituting \eqref{ccx3}, \eqref{cv3} and \eqref{fd5}  into \eqref{cv2} and by using \eqref{kd2} and \eqref{kd1}, the BER of the considered system under no knowledge of $h$ is derived in \eqref{cn5}.

\bibliographystyle{IEEEtran}
%\vspace{-2 mm}
%\balance
%\bibitem[$\rm R2.17$]{FSO_survey}
%M. A. Khalighi and Murat Uysal, ``Survey on free space optical communication: A Communication theory perspective,'' \emph{IEEE Commun. Survey \& Tutorials}, vol. 16. no. 4, pp. 2231-2258, Third Quarter, 2014.
%\bibliography{IEEEabrv,myref}

\end{document}